\documentclass[british,nofootinbib, 11pt,a4paper]{revtex4-1}
\usepackage[T1]{fontenc}
\usepackage[latin9]{inputenc}
\usepackage{mathtools}
\usepackage{amsmath}
\usepackage{amsthm}
\usepackage{amssymb}

\makeatletter
\theoremstyle{remark}
\newtheorem*{rem*}{\protect\remarkname}

\makeatother

\usepackage{babel}
\providecommand{\remarkname}{Remark}

\begin{document}

\title{A Holographic Construction of MHV Graviton Amplitudes in Celestial
CFT}

\author{Igor Mol}

\affiliation{State University of Campinas (Unicamp)}
\email{igormol@ime.unicamp.br}

\selectlanguage{british}%
\begin{abstract}
In this note, we show that tree-level MHV $n$-graviton amplitudes,
when analytically continued to Klein space, can be holographically
generated as the correlators of a two-dimensional conformal field
theory ($2d$ CFT) in the large-$N$ semiclassical $b\rightarrow0$
limit. This $2d$ CFT is composed of a Liouville field theory, coupled
to a level-one $SO\left(N\right)$ Kac-Moody current, and a doublet
of free fermions residing on $\mathbf{CP}^{1}$.
\end{abstract}
\maketitle

\section{Introduction}

An important problem in the bottom-up approach to celestial holography
is the formulation of simplified celestial conformal field theories
(CFTs) that captures basic characteristics of quantum gravity in flat
spacetime. Significant progress in this direction was achieved by
\citet{melton2024celestial}, who introduced a two-dimensional CFT
($2d$ CFT) on the celestial sphere consisting of a Liouville field
theory, a level-one $SO\left(N\right)$ Kac-Moody current algebra,
and a free fermion. Remarkably, in the large-$N$ semiclassical limit,
the correlation functions of this CFT successfully reproduce tree-level
MHV gluon scattering amplitudes. This work follows a sequence of studies
by \citet{casali2022celestial} and \citet{melton2024celestial},
arising from the observation that the translation invariance inherent
in $4d$ scattering amplitudes of massless particles (expressed as
distributions arising from energy-momentum conservation) fails to
manifest within the correlation functions of conventional $2d$ CFTs.

To address this issue, \citet{stieberger2023yang} proposed an approach
that considers the possibility of breaking translation invariance
by analysing Yang-Mills theory on a spherical dilaton shockwave background.
This led to the observation that tree-level MHV gluon amplitudes can
be reformulated as correlation functions of a Liouville field theory.
Parallel to this, \citet{casali2022celestial}, followed by \citet{melton2023celestial},
initiated an alternative program, based on the decomposition of celestial
amplitudes as weighted integrals of $AdS_{3}$-Witten diagrams, defined
on the leaves of a hyperbolic foliation of Klein spacetime (cf. \citet{barrett1994kleinian,bhattacharjee2022celestial}
for mathematical details). In a manner characteristic of celestial
holography, where distinct research lines often converge in unanticipated
ways, \citet{melton2024celestial} employed insights from both the
celestial leaf amplitude program and the approach due to \citet{stieberger2023yang,stieberger2023celestial}
for addressing the distributional nature of massless scattering amplitudes.
This led to an important conclusion: by revisiting the work of \citet{nair1988current,nair2005note}
and combining it with recent mathematical results by \citet{costello2022celestial}
as well as \citet{bu20224d}, it became possible to formulate a simple
model for celestial CFT. As explained in detail by \citet{melton2024celestial},
the correlation functions of this $2d$ CFT, in the large-$N$ semiclassical
limit, can be employed to express the tree-level MHV gluon amplitudes.

Inspired by this significant result and motivated by the goal of extending
our understanding of the holographic principle to flat space, we recognised
the necessity of investigating how the work of \citet{melton2024celestial}
could be extended to generate graviton scattering amplitudes. Our
belief that a $2d$ CFT on the celestial sphere could reproduce (in
a certain limit) the tree-level scattering amplitudes of the MHV subsector
of Einstein's gravity was also influenced by the works of \citet{pate2021celestial,banerjee2021mhv,guevara2021holographic,himwich2022celestial,strominger2021w,banerjee2024all}.
These authors showed that the celestial operator product expansions
(OPEs) of conformal primary gluons and gravitons are constrained by
the collinear singularities of gauge and gravity amplitudes as well
as by asymptotic symmetries. Moreover, the current algebra studied
by \citet{banerjee2021mhvx}, along with the system of linear first-order
partial differential equations satisfied by MHV $n$-graviton amplitudes
(analogous to the \citet{knizhnik1984current} equations for the Wess-Zumino-Witten
model) provides compelling evidence for the existence of an autonomous
sector within celestial CFT which holographically encode the MHV graviton
amplitudes, functioning as a ``minimal model'' for celestial holography. 

The organisation of our paper is as follows. In Section 2, we discuss
in considerable detail our selection of the matter system comprising
the $2d$ CFT which generates the tree-level MHV \emph{$4$}-graviton
amplitude. This approach serves a dual purpose: it not only provides
a physical motivation for the selection of our $2d$ model but also
renders the more abstract and general arguments for the construction
of the $n$-point amplitude in Section 3 more accessible. In the concluding
section, following a concise review of our principal findings, we
discuss several unresolved issues that, we hope, will inspire further
investigation into the development of toy models for celestial holography.

\section{Model Building: The Case of the Four-Point Function}

To motivate the formulation of a two-dimensional conformal field theory
($2d$ CFT) on the celestial sphere, which we propose as the holographic
dual to the MHV subsector of Einstein's gravity in flat spacetime,
we begin by analysing the tree-level MHV\footnote{Henceforth, all graviton amplitudes discussed shall be understood
as tree-level MHV, which obviates the need for repetitive clarification.} $4$-graviton amplitude $M_{4}$ for the scattering of gravitons
$1^{-}$, $2^{-}$, $3^{+}$ and $3^{+}$. In the course of this analysis,
we will introduce the corresponding matter fields on the complex projective
line $\mathbf{CP}^{1}$ (which we identify with the celestial sphere
at null-infinity) which are constructed to reproduce the graviton
amplitudes in a certain limit. This ensures that the generalisation
of our approach is both mathematically rigorous and physically motivated.
Once the specific method by which the $4$-graviton amplitude is expressed
as a correlation function of a family of operators within a $2d$
CFT is thoroughly understood, the derivation of the $3$-point function
becomes obvious, while the extension to $n$-graviton amplitudes (with
$n\geq5$) arises naturally from our proposed CFT.

Let $\hat{\chi}\left(z\right)$ and $\hat{\chi}^{\dagger}\left(z\right)$
be a doublet of fermions on $\mathbf{CP}^{1}$ with mode expansions:
\begin{equation}
\hat{\chi}\left(z\right)\coloneqq\sum_{n=0}^{\infty}b_{n}z^{n},\,\,\,\hat{\chi}^{\dagger}\left(z\right)\coloneqq\sum_{n=0}^{\infty}b_{n}^{\dagger}z^{-n-1},\label{eq:Mode}
\end{equation}
where $\{b_{m},b_{n}^{\dagger}\}=\delta_{mn}$ are fermionic creation
and annihilation operators acting on a vacuum state $|0\rangle$,
such that $b_{m}|0\rangle=0$. Let $u^{A}=\left(\xi,\zeta\right)^{T}\in\mathbf{C}^{2}$
be a two-component spinor with $\zeta\neq0$, where $z=\xi/\zeta\in\mathbf{CP}^{1}$
represents its local coordinate on the complex projective line. We
define the pair of fermions $\chi\left(u\right)$ and $\chi^{\dagger}\left(u\right)$
as $\chi\left(u\right)\coloneqq\zeta^{-1}\hat{\chi}\left(z\right)$
and $\chi^{\dagger}\left(u\right)\coloneqq\zeta^{-1}\hat{\chi}^{\dagger}\left(z\right)$.
Observe that while the doublet $\left(\hat{\chi},\hat{\chi}^{\dagger}\right)$
represents free fermions residing on $\mathbf{CP}^{1}$ the doublet
$\left(\chi,\chi^{\dagger}\right)$ corresponds to fields in the space
of two-component spinors.

Let $\lambda^{A}$ be an auxiliary two-component spinor, $x^{\mu}$
a spacetime four-vector and $p^{\mu}=\omega q^{\mu}$ a null four-vector
representing the momentum of a graviton. Here, $q^{\mu}\coloneqq(\sigma^{\mu})_{A\dot{A}}v^{A}\bar{v}^{\dot{A}}$
is our convention for the standard null momentum, and utilising the
little group rescaling property, we choose local coordinates on the
complex projective line such that\footnote{Note that our convention for normalising the spinor-helicity variables,
which parametrise the standard null momentum, differs from that adopted
in the works of \citet{pasterski2017gluon,pasterski2017conformal,cheung20174d}.
In those references, the conventions are such that $v^{A}=\sqrt[]{2}\left(1,\bar{z}\right)$
and $\bar{v}^{\dot{A}}=-\sqrt[]{2}\left(1,z\right)$.} $v^{A}=\left(z,1\right)^{T}$ and $\bar{v}^{\dot{A}}=\left(\bar{z},1\right)^{T}$.
Then, we define the differential operators $\mathcal{Q}\left(u,\bar{u}\right)$
and \textbf{$\mathcal{P}\left(u,\bar{u}\right)$} as follows:
\begin{equation}
\mathcal{Q}\left(u,\bar{u}\right)\coloneqq\chi^{\dagger}\left(u\right)\chi\left(u\right)e^{ipx},\,\,\,\mathcal{P}\left(u,\bar{u}\right)\coloneqq-i\frac{\lambda^{A}\bar{u}^{\dot{A}}}{\left\langle u\lambda\right\rangle }\frac{\partial}{\partial x^{A\dot{A}}}e^{ipx},\label{eq:Q-P-Operators}
\end{equation}
For brevity, we denote the Minkowski inner product as a juxtaposition,
$px\coloneqq p\cdot x$, and the integral measure as $\int\left[dxd\lambda\right]\coloneqq\int\frac{d^{4}x}{\left(2\pi\right)^{4}}\int\frac{dz_{\lambda}}{\pi}$.
The spinor derivative is denoted by $\bar{\partial}\coloneqq\partial/\partial\bar{\lambda}$.
Furthermore, let $u_{1}^{A}$, $u_{2}^{A}$, ... $\in\mathbf{C}^{2}$
and $\bar{u}_{1}^{\dot{A}}$, $\bar{u}_{2}^{\dot{A}}$, ... $\in\mathbf{C}^{2}$
be sequences of two-component spinors. We denote the operators $\mathcal{Q}$
and $\mathcal{P}$ evaluated at $\left(u_{k},\bar{u}_{k}\right)$
as $\mathcal{Q}_{k}\coloneqq\mathcal{Q}\left(u_{k},\bar{u}_{k}\right)$
and $\mathcal{P}_{k}\coloneqq\mathcal{P}\left(u_{k},\bar{u}_{k}\right)$.
The cyclic permutation of the contractions of all spinors $u_{k}$
is $C\left(n\right)\coloneqq\left\langle 12\right\rangle \left\langle 23\right\rangle ...\left\langle n1\right\rangle $.

In Appendix \ref{subsec:M3-M4-Amplitudes} (cf. Eq. (\ref{eq:M3-final})),
we have demonstrated that the $4$-graviton amplitude can be expressed
as:
\begin{equation}
M_{4}=-\frac{\omega_{1}^{3}\omega_{2}^{3}}{\omega_{3}\omega_{4}}\frac{z_{12}^{8}}{z_{12}z_{23}z_{34}z_{41}}\int\left[dxd\lambda\right]\,\langle\lambda\big|\mathcal{Q}_{1}\mathcal{P}_{2}\mathcal{Q}_{3}\mathcal{Q}_{4}\bar{\partial}\big|\lambda\rangle.\label{eq:Four-Graviton}
\end{equation}

In accordance with the celestial holography dictionary (as explained
in \citet{pasterski2021lectures,pasterski2023chapter,raclariu2021lectures,strominger2018lectures,aneesh2022celestial,donnay2024celestial,puhmcelestial}),
the \emph{celestial $4$-graviton amplitude} $\widehat{M}_{4}$ corresponding
to $M_{4}$ is the $\epsilon$-regulated Mellin transform:
\begin{equation}
\widehat{M}_{4}\left(\left\{ \Delta_{i},z_{i},\bar{z}_{i}\right\} \right)=\prod_{k=1}^{4}\int_{0}^{\infty}d\omega_{k}\,\,\,\omega_{k}^{\Delta_{k}-1}e^{-\epsilon\omega_{k}}M_{4}\left(\left\{ \omega_{i},z_{i},\bar{z}_{i}\right\} \right).\label{eq:Celestial-Amplitude}
\end{equation}

For any integer $\alpha$ and $\Delta\in1+i\mathbf{R}$, let $\hat{\mathcal{Q}}_{\alpha,\Delta}\left(z,\bar{z}\right)$
and $\hat{\mathcal{P}}_{\alpha,\Delta}\left(z,\bar{z}\right)$ represent
the $\epsilon$-regulated Mellin transforms of the operators $\mathcal{Q}\left(u,\bar{u}\right)$
and $\mathcal{P}\left(u,\bar{u}\right)$, respectively, \emph{each
weighted by powers of the frequency} $\omega^{\alpha}$. Explicitly,
these transforms are given by\footnote{To compute the Mellin transform of $\mathcal{Q}\left(u,\bar{u}\right)$,
it is first necessary to replace the fermions $\chi\left(u\right)$
and $\chi^{\dagger}\left(u\right)$, which are fields defined on the
space of two-component spinors, with the pair $\hat{\chi}\left(z\right)$
and $\hat{\chi}^{\dagger}\left(z\right)$ defined on $\mathbf{CP}^{1}$.
This substitution is given by: 
\begin{equation}
\chi\left(u\right)\coloneqq\zeta^{-1}\hat{\chi}\left(z\right),\,\,\,\chi^{\dagger}\left(u\right)\coloneqq\zeta^{-1}\hat{\chi}^{\dagger}\left(z\right).
\end{equation}
 Recalling that, in our local coordinate $z\in\mathbf{CP}^{1}$, we
parametrise $u^{A}$ by $\sqrt[]{\omega}\left(z,1\right)^{T}$, we
have:
\begin{equation}
\mathcal{Q}\left(u,\bar{u}\right)=\frac{e^{i\omega qx}}{\omega}\hat{\chi}^{\dagger}\left(z\right)\hat{\chi}\left(z\right).\label{eq:Q-new}
\end{equation}
Since $\hat{\chi}\left(z\right)$ and $\hat{\chi}^{\dagger}\left(z\right)$
are independent of the frequency $\omega$, the Mellin transform of
$\mathcal{Q}\left(u,\bar{u}\right)$ follows straightforwardly from
this expression.}:
\begin{equation}
\widehat{\mathcal{Q}}_{\alpha,\Delta}\left(u,\bar{u}\right)\coloneqq\int_{0}^{\infty}d\omega\,\,\,\omega^{\Delta-1}e^{-\epsilon\omega}\omega^{\alpha}\mathcal{Q}\left(u,\bar{u}\right)=\frac{\Gamma\left(\Delta+\alpha-1\right)}{\left(\epsilon-iqx\right)^{\Delta+\alpha-1}}\hat{\chi}^{\dagger}\left(z\right)\hat{\chi}\left(z\right),\label{eq:Mellin-transformed-operator-1}
\end{equation}
\begin{equation}
\widehat{\mathcal{P}}_{\alpha,\Delta}\left(u,\bar{u}\right)\coloneqq\int_{0}^{\infty}d\omega\,\,\,\omega^{\Delta-1}e^{-\epsilon\omega}\mathcal{P}_{\lambda}\left(u,\bar{u}\right)=-i\frac{\lambda^{A}\bar{u}^{\dot{A}}}{\left\langle u\lambda\right\rangle }\frac{\partial}{\partial x^{A\dot{A}}}\frac{\Gamma\left(\Delta+\alpha\right)}{\left(\epsilon-iqx\right)^{\Delta+\alpha}},\label{eq:Mellin-transformed-operator-2}
\end{equation}
To simplify our notation, and where no confusion will arise, we shall
denote these operators by $\widehat{\mathcal{Q}}_{k}\coloneqq\widehat{\mathcal{Q}}_{\alpha_{k},\Delta_{k}}\left(u_{k},\bar{u}_{k}\right)$
and $\widehat{\mathcal{P}}_{k}\coloneqq\widehat{\mathcal{P}}_{\alpha_{k},\Delta_{k}}\left(u_{k},\bar{u}_{k}\right)$.
With this convention, and by setting $\alpha_{1}=\alpha_{2}=3$ and
$\alpha_{3}=\alpha_{4}=-1$, which are powers of the frequencies $\omega_{k}$
$\left(k=1,2,...\right)$ that factor into the first term of Eq. (\ref{eq:Four-Graviton}),
it follows from Eqs. (\ref{eq:Four-Graviton}, \ref{eq:Celestial-Amplitude})
that the celestial $4$-graviton amplitude can be written as:
\begin{equation}
\widehat{M}_{4}=-\frac{z_{12}^{8}}{z_{12}z_{23}z_{34}z_{41}}\int\left[dxd\lambda\right]\langle\lambda\big|\widehat{\mathcal{Q}}_{1}\widehat{\mathcal{P}}_{2}\widehat{\mathcal{Q}}_{3}\widehat{\mathcal{Q}}_{4}\bar{\partial}\big|\lambda\rangle.
\end{equation}

In Appendix \ref{sec:Operator-Factorisation}, we proved the following
result. Let $x^{\mu}$, $y^{\mu}$, $q^{\mu}$, $q_{1}^{\mu}$, $q_{2}^{\mu}$,
... be a sequence of four-vectors, and $\varepsilon$, $\varepsilon_{1}$,
$\varepsilon_{2}$, ... a family of infinitesimal variables. Let $G_{\Delta}\left(x,q\right)$
and $K_{k,\Delta}\left(x,q\right)$ ($k=1,2,...$) be the functions
given by:
\begin{equation}
G_{\Delta}\left(x,q\right)\coloneqq\frac{\Gamma\left(\Delta\right)}{\left(\varepsilon-iqx\right)^{\Delta}},\,\,\,K_{k,\Delta}\left(x,q\right)\coloneqq\frac{\Gamma\left(\Delta\right)}{\left(\varepsilon_{k}-iqx\right)^{\Delta}}.\label{eq:Functions}
\end{equation}
Here, $\Delta$ will be referred to as the ``conformal weight,'' in
anticipation of its significance for our CFT. To simplify notation
where no ambiguity arises, we shall write $G_{k}=G_{\Delta}\left(x,q_{k}\right)$
and $K_{k,\Delta}=K_{k,\Delta}\left(x,q_{k}\right)$. It is important
to note that the main distinction between $G_{\Delta}\left(x,q\right)$
and $K_{k,\Delta}\left(x,q\right)$ is that the ``infinitesimal regulators''
$\varepsilon_{k}$ associated with the functions $\{K_{k,\Delta}\left(x,q\right)\}$
are mutually independent, whereas the family of functions $\{G_{\Delta}\left(x,q\right)\}$
is regulated by a common parameter $\varepsilon$. The form of these
functions should already be familiar to the reader, as they correspond,
when restricted to the de Boer-Solodukhin hyperbolic foliation of
Minkowski space (cf. \citet{de2003holographic}), to the $AdS_{3}$-Witten
bulk-to-boundary propagator, modulo the normalisation constants of
the Green function of the Laplacian $\square_{AdS}$. (For a review,
see \citet{penedones2017tasi}.) 

Let $\Delta_{1}$, $\Delta_{2}$, ... $\in1+i\mathbf{R}$ be a sequence
of conformal weights. We define the differential operators $\mathbf{X}_{k}\left(x\right)$,
$\mathcal{D}_{k}\left(y\right)$ and $\mathcal{T}_{k}\left(y\right)$,
which act on the space of functions $\{G_{\Delta}\left(x,q\right),K_{k,\Delta}\left(x,q\right)\}$,
as follows:
\begin{equation}
\mathbf{X}_{k}\left(x\right)\coloneqq M_{k}^{A\dot{A}}\frac{\partial}{\partial x^{A\dot{A}}}\frac{\Gamma\left(\Delta_{k}\right)}{\left(\varepsilon-iq_{k}\cdot x\right)^{\Delta_{k}}},\label{eq:Operator-X}
\end{equation}
\begin{equation}
\mathcal{D}_{k}\left(y\right)\coloneqq M_{k}^{A\dot{A}}\frac{\partial}{\partial y^{A\dot{A}}}e^{-iq_{k}\cdot y\partial_{\varepsilon_{k}}},\,\,\,\mathcal{T}_{k}\left(y\right)\coloneqq e^{-iq_{k}\cdot y\partial_{\varepsilon_{k}}},\label{eq:Operators-DT}
\end{equation}
where $M_{k}^{A\dot{A}}$ is a matrix dependent on the spinors $\lambda^{A}$
and $\bar{u}_{k}^{\dot{A}}$ introduced earlier, and $\partial_{\varepsilon_{k}}\coloneqq\partial/\partial\varepsilon_{k}$.
Then, according to the Proposition proved in Subsec. \ref{subsec:Proof-of-Eq.}
(cf. Eq. (\ref{eq:Theorem})),
\begin{equation}
\mathbf{X}_{1}\mathbf{X}_{2}...\mathbf{X}_{n}G_{n+1}G_{n+2}=\int\left[dy\right]\,\delta_{y}\mathcal{D}_{1}\mathcal{D}_{2}...\mathcal{D}_{n}\mathcal{T}_{n+1}\mathcal{T}_{n+2}\prod_{r=1}^{n+2}K_{r,\Delta_{r}}\left(x\right),\label{eq:Theorem-1}
\end{equation}
where, for simplicity, we will write $\int\left[dy\right]\delta_{y}\coloneqq\int d^{4}y\delta^{\left(4\right)}\left(y\right)$
and omit the variable $y$ on $\mathcal{D}_{k}$ and $\mathcal{T}_{k}$.
Now, note that from Eqs. (\ref{eq:Mellin-transformed-operator-1},
\ref{eq:Mellin-transformed-operator-2}), the Mellin-transformed operators
$\hat{\mathcal{Q}}_{\ell}$ and $\hat{\mathcal{P}}_{\ell}$ can be
expressed in terms of $\mathbf{X}_{k}\left(x\right)$ and $G_{\Delta}\left(x,q\right)$
as:
\begin{equation}
\hat{\mathcal{P}}_{k}=\mathbf{X}_{k},\,\,\,\text{and}\;\hat{\mathcal{Q}}_{k}=G_{\Delta_{k}+\alpha_{k}-1}\left(x,q_{k}\right)\hat{\chi}^{\dagger}\left(z_{k}\right)\hat{\chi}\left(z_{k}\right),
\end{equation}
provided that we take $M_{\ell}^{A\dot{A}}=-i\lambda^{A}\bar{u}^{\dot{A}}/\left\langle \ell\lambda\right\rangle .$
Hence, introducing a new operator $\mathcal{R}_{k}$ by: 
\begin{equation}
\mathcal{R}_{k,y}\coloneqq\mathcal{T}_{k}\left(y\right)\hat{\chi}^{\dagger}\left(z_{k}\right)\hat{\chi}\left(z_{k}\right),\label{eq:Operator-R}
\end{equation}
Eq. (\ref{eq:Theorem-1}) implies that\footnote{We omitted the dependence on $y$ to streamline the notation.}:
\begin{equation}
\widehat{M}_{4}=-\frac{\left\langle 12\right\rangle ^{8}}{C\left(4\right)}\int\left[dydz_{\lambda}\right]\delta_{y}\langle\lambda\big|\mathcal{R}_{1}\mathcal{D}_{2}\mathcal{R}_{3}\mathcal{R}_{4}\bar{\partial}\big|\lambda\rangle\int\left[dx\right]\prod_{k=1}^{4}\frac{\Gamma\left(\beta_{k}\right)}{\left(\varepsilon_{k}-iq_{k}x\right)^{\beta_{k}}},\label{eq:Argument-1}
\end{equation}
where: 
\begin{equation}
\beta_{k}=\Delta_{k}+\alpha_{k}-1\,\left(k=1,3,4\right),\,\,\,\beta_{2}=\Delta_{2}+\alpha_{2}.
\end{equation}

To evaluate the final integral on the right-hand side of Eq. (\ref{eq:Argument-1}),
we substitute the infinitesimal parameters $\varepsilon_{k}$ with
$\epsilon\varepsilon_{k}$, where $\epsilon\rightarrow0$, and treat
$\varepsilon_{k}$ as (non-infinitesimal) real and independent variables.
Consequently, the integrand becomes:
\begin{equation}
\prod_{k=1}^{4}\frac{\Gamma\left(\beta_{k}\right)}{\left(\varepsilon_{k}-iq_{k}x\right)^{\beta_{k}}}=\prod_{k=1}^{4}\frac{e^{-\beta_{k}\log\varepsilon_{k}}\Gamma\left(\beta_{k}\right)}{\left(\epsilon-i\ell_{k}x/\varepsilon_{k}\right)^{\beta_{k}}}.
\end{equation}
Here, $\ell_{k}^{\mu}$ is the four-vector $\ell_{k}^{\mu}\coloneqq(\sigma^{\mu})_{A\dot{A}}\pi_{k}^{A}\bar{\pi}_{k}^{\dot{A}}$,
where the two-component spinors $\pi_{k}^{A}$ and $\bar{\pi}_{k}^{\dot{A}}$
are parametrised in local coordinates $z_{k},\bar{z}_{k}$ $\in\mathbf{CP}^{1}$
by $\pi_{k}=\varepsilon_{k}^{-1/2}\left(z_{k},1\right)^{T}$ and $\bar{\pi}_{k}=\varepsilon_{k}^{-1/2}\left(\bar{z}_{k},1\right)^{T}$.
Utilising the results developed in Appendix A, we find, in the semiclassical
$b\rightarrow0$ limit of Liouville field theory $\hat{\phi}\left(z,\bar{z}\right)$
residing on $\mathbf{S}^{2}$, that:
\begin{align}
 & \int\left[dx\right]\prod_{k=1}^{4}\frac{\Gamma\left(\beta_{k}\right)}{\left(\varepsilon_{k}-iq_{k}x\right)^{\beta_{k}}}=\delta_{\beta}\left\langle \prod_{k=1}^{4}e^{-\beta_{k}\log\varepsilon_{k}}\Gamma\left(\beta_{k}\right)V_{b\beta_{k}}\left(\frac{z_{k}}{\sqrt{\varepsilon_{k}}},\frac{\bar{z}_{k}}{\sqrt{\varepsilon_{k}}}\right)\right\rangle \\
 & +\left(\bar{z}\leftrightarrow-\bar{z}\right),\label{eq:Argument-2}
\end{align}
where $\left(\bar{z}\leftrightarrow-\bar{z}\right)$ signifies the
replication of the first term with the substitution of $\bar{z}$
by $-\bar{z}$. We denote by $V_{\alpha}\left(z,\bar{z}\right)=:\exp(2\alpha\hat{\phi}\left(z,\bar{z}\right)):$
the Liouville vertex operator with momentum $\alpha$. For the sake
of simplicity, we have defined the normalised delta function associated
with the conformal weights $\beta_{k}=\beta\left(\Delta_{k},\alpha_{k}\right)$
as:
\begin{equation}
\delta_{\beta}\coloneqq\frac{1}{8\pi^{3}\mathcal{N}_{\mu,b}}\delta\left(4-\sum_{r=1}^{4}\beta_{r}\right),\label{eq:Delta-Conformal}
\end{equation}
where $\mathcal{N}_{\mu,b}$ is a constant defined in Eq. (\ref{eq:Constant})
depending on the Liouville ``cosmological constant.'' 

Following the seminal work of \citet{nair1988current}, and in the
context of celestial holography, the works by \citet{costello2022celestial,melton2024celestial,bu20224d,stieberger2023celestial,stieberger2023yang},
the Parke-Taylor denominator---ubiquitous in gauge theory amplitudes---can
be reproduced using a matter system consisting of chiral fermionic
primary free fields $\psi^{i}$ on $\mathbf{CP}^{1}$ obeying the
operator product expansions (OPEs):
\begin{equation}
\psi^{i}\left(u_{k}\right)\psi^{j}\left(u_{\ell}\right)\sim\frac{\delta^{ij}}{z_{k\ell}},\label{eq:Matter-OPE}
\end{equation}
where $z_{k\ell}\coloneqq z_{k}-z_{\ell}$. These fields, transforming
under the adjoint representation of the group $SO\left(N\right)$,
give rise to level-one Kac-Moody currents:
\begin{equation}
J^{a}\left(z\right)\coloneqq\frac{1}{2}R_{ij}^{a}:\psi^{i}\psi^{j}:\left(z\right),\label{eq:Currents}
\end{equation}
whose OPEs are given by:
\begin{equation}
J^{a}\left(z_{k}\right)J^{b}\left(z_{\ell}\right)\sim\frac{1}{z_{k\ell}^{2}}\delta^{ab}+\frac{1}{z_{k\ell}}f^{abc}J^{c}\left(u_{\ell}\right).\label{eq:OPE}
\end{equation}
In the construction of these currents, Einstein's summation convention
has been employed. Here, $a,b,c,$ etc., represent indices corresponding
to the adjoint representation of $SO\left(N\right)$ wherein $\left[R^{a},R^{b}\right]=f^{abc}R^{c}$,
with generators $R^{a}$ normalised such that $\mathbf{Tr}\left(R^{a}R^{b}\right)=2\delta^{ab}$
and structure constants denoted by the symbols $f^{abc}$.

The $SO\left(N\right)$ symmetry that governs the currents, together
with the invariance under $SL\left(2,\mathbf{C}\right)$ transformations,
$z\mapsto z'$, implies (see \citet{ketov1995conformal}) that the
leading term of the correlation functions constructed from $J^{a}$
in the \emph{large-$N$ limit} is dominated by:
\begin{equation}
\left\langle J^{a_{1}}\left(z_{1}\right)J^{a_{2}}\left(z_{2}\right)...J^{a_{n}}\left(z_{n}\right)\right\rangle \underset{N\gg1}{\sim}\mathbf{Tr}\left(R^{a_{1}}R^{a_{2}}...R^{a_{n}}\right)\frac{1}{\left\langle 12\right\rangle \left\langle 23\right\rangle ...\left\langle n1\right\rangle }.
\end{equation}
Henceforth, we define $k_{n}\coloneqq\mathbf{Tr}\left(R^{1}R^{2}...R^{n}\right)$,
and proceed by omitting the colour indices, and write $J_{k}\coloneqq J\left(z_{k}\right)$,
allowing us to succinctly express:
\begin{equation}
\left\langle \prod_{k=1}^{n}J_{k}\right\rangle =\frac{1}{z_{12}z_{23}...z_{n1}}.\label{eq:Correlator}
\end{equation}
Consequently, the cyclic term appearing in Eq. (\ref{eq:Four-Graviton})
becomes: 
\begin{equation}
C\left(4\right)=\left\langle J_{1}J_{2}J_{3}J_{4}\right\rangle .\label{eq:Argument-3-1}
\end{equation}

The final phase of our argument is to derive a field-theoretic representation
for the term $z_{12}^{8}$ within the celestial $4$-graviton amplitude
$\widehat{M}_{4}$. This can be achieved through the application of
Berezin calculus (see \citet{berezin2013introduction}). Let $\theta_{A}^{p}$
$\left(p=1,2,...,8\right)$ be Grassmann-valued two-component spinors
satisfying the normalisation condition $\int d^{2}\theta^{p}\,\theta_{A}^{p}\theta_{B}^{q}=\varepsilon_{AB}$,
for each $p=1,2,...,8$, and define the two-component spinors $v_{1}^{A}=\left(z_{1},1\right)^{T}$
and $v_{2}^{A}=\left(z_{2},1\right)^{T}$. Therefore, 
\begin{equation}
z_{12}^{8}=\int d^{16}\theta\prod_{p=1}^{8}\theta_{A}^{p}v_{1}^{A}\prod_{q=1}^{8}\theta_{B}^{q}v_{2}^{q}.\label{eq:Argument-3}
\end{equation}
To simplify our notation for the integration measure, and to render
it analogous to the integrals over the measure $\left[dyd\lambda\right]$,
we introduce the compact notation: $\int\left[d\theta\right]\coloneqq\int d^{16}\theta\prod_{p=1}^{8}\theta_{A}^{p}v_{i}^{A}\prod_{q=1}^{8}\theta_{B}^{q}v_{j}^{B}$.

From Eqs. (\ref{eq:Argument-1}, \ref{eq:Argument-2}, \ref{eq:Argument-3}),
the four-graviton amplitude admits the following integral representation:
\begin{align}
\widehat{M}_{4} & =-\delta_{\beta}\int\left[d\theta dyd\lambda\right]\delta_{y}\left\langle J_{1}J_{2}J_{3}J_{4}\right\rangle \langle\lambda\big|\mathcal{R}_{1}\mathcal{D}_{2}\mathcal{R}_{3}\mathcal{R}_{4}\bar{\partial}\big|\lambda\rangle\\
 & \left\langle \prod_{k=1}^{4}e^{-\beta_{k}\log\varepsilon_{k}}\Gamma\left(\beta_{k}\right)V_{b\beta_{k}}\left(\frac{z_{k}}{\sqrt{\varepsilon_{k}}},\frac{\bar{z}_{k}}{\sqrt{\varepsilon_{k}}}\right)\right\rangle +\left(\bar{z}\leftrightarrow-\bar{z}\right).
\end{align}
Finally, let $\mathcal{G}_{\varepsilon,y,\beta}\left(z,\bar{z}\right)$
and $\mathcal{H}_{\varepsilon,y,\beta}\left(z,\bar{z}\right)$ be
the $\mathbf{CP}^{1}$ operators with conformal weight $\beta$ defined
by:
\begin{equation}
\mathcal{G}_{\varepsilon,y,\beta}\left(z,\bar{z}\right)\coloneqq\Gamma\left(\beta\right)J\left(z\right)\mathcal{R}_{y}\left(z,\bar{z}\right)e^{-\beta\log\varepsilon}V_{b\beta}\left(\frac{z}{\sqrt[]{\varepsilon}},\frac{\bar{z}}{\sqrt[]{\varepsilon}}\right),\label{eq:Operator-G}
\end{equation}
\begin{equation}
\mathcal{H}_{\varepsilon,y,\beta}\left(z,\bar{z}\right)\coloneqq\Gamma\left(\beta\right)J\left(z\right)\mathcal{D}_{y}\left(z,\bar{z}\right)e^{-\beta\log\varepsilon}V_{b\beta}\left(\frac{z}{\sqrt{\varepsilon}},\frac{\bar{z}}{\sqrt{\varepsilon}}\right).\label{eq:Operator-H}
\end{equation}

Therefore, the celestial tree-level MHV $4$-graviton amplitude can
be written as the $\mathbf{CP}^{1}$ correlation function:
\begin{align}
\widehat{M}_{4} & =-\delta_{\beta}\int\left[d\theta dyd\lambda\right]\delta_{y}\langle\mathcal{G}_{\varepsilon_{1},y,\beta_{1}}\left(z_{1},\bar{z}_{1}\right)\mathcal{H}_{\varepsilon_{2},y,\beta_{2}}\left(z_{2},\bar{z}_{2}\right)\label{eq:Four-Graviton-1}\\
 & \mathcal{G}_{\varepsilon_{3},y,\beta_{3}}\left(z_{3},\bar{z}_{3}\right)\mathcal{G}_{\varepsilon_{4},y,\beta_{4}}\left(z_{4},\bar{z}_{4}\right)\rangle+\left(\bar{z}\leftrightarrow-\bar{z}\right).
\end{align}

\begin{rem*}
The integral over $dy$ in Eq. (\ref{eq:Four-Graviton-1}) should
not be regarded as a spacetime integral, as it is localised around
$y=0$ via the delta function $\delta_{y}=\delta^{\left(4\right)}\left(y\right)$.
The four-vector $y^{\mu}$ should instead be understood as a \emph{continuous
label} of the operators, in conjunction with the conformal weights
$\beta_{1}$, $\beta_{2}$, etc. Furthermore, the order of the factors
in Eqs. (\ref{eq:Operator-G}, \ref{eq:Operator-H}) is important,
as $\mathcal{R}_{y}\left(z,\bar{z}\right)$ and $\mathcal{D}_{y}\left(z,\bar{z}\right)$
contain the operator $e^{-iqx\partial_{\varepsilon}}$, which acts
on the continuous parameter $\varepsilon$.
\end{rem*}
To conclude this section, we observe that the celestial $3$-graviton
amplitude can now be formulated as a correlator of $\mathbf{CP}^{1}$
fields. As demonstrated in Appendix \ref{subsec:Proof-of-Eq.}, the
$3$-graviton amplitude $M_{3}$ can be expressed as:
\begin{equation}
M_{3}=-\frac{\omega_{1}^{3}\omega_{2}^{3}}{\omega_{3}}\frac{z_{12}^{8}}{z_{12}z_{23}z_{31}}\int\left[dxd\lambda\right]\,\langle\lambda\big|\mathcal{Q}_{1}\mathcal{Q}_{2}\mathcal{Q}_{3}\bar{\partial}\big|\lambda\rangle.
\end{equation}
Here, we define $\alpha_{1}=\alpha_{2}=3$ and $\alpha_{3}=-1$, and
let $\beta_{k}=\Delta_{k}+\alpha_{k}-1$ for $1\leq k\leq3$. The
corresponding celestial amplitude $\widehat{M}_{3}$ takes the form:
\begin{equation}
\widehat{M}_{3}=-\frac{z_{12}^{8}}{z_{12}z_{23}z_{31}}\int\left[dyd\lambda\right]\delta_{y}\langle\lambda\big|\mathcal{R}_{1}\mathcal{R}_{2}\mathcal{R}_{3}\bar{\partial}\big|\lambda\rangle\int\left[dx\right]\prod_{k=1}^{3}\frac{\Gamma\left(\beta_{k}\right)}{\left(\varepsilon_{k}-iq_{k}x\right)^{\beta_{k}}}.
\end{equation}
Recall that the operators $\mathcal{R}_{y}\left(z,\bar{z}\right)$
were introduced in Eq. (\ref{eq:Operator-R}). Finally, by solving
the integral over the measure $\left[dx\right]$ using the techniques
outlined in Appendix A, and subsequently reformulating the resulting
expression in terms of the operators $\mathcal{G}_{\varepsilon,y,\beta}\left(z,\bar{z}\right)$
as defined in Eq. (\ref{eq:Operator-G}), we conclude that:
\begin{equation}
\widehat{M}_{3}=-\delta_{\beta}\int\left[d\theta dyd\lambda\right]\delta_{y}\langle\mathcal{G}_{\varepsilon_{1},y,\beta_{1}}\left(z_{1},\bar{z}_{1}\right)\mathcal{G}_{\varepsilon_{2},y,\beta_{2}}\left(z_{2},\bar{z}_{2}\right)\mathcal{G}_{\varepsilon_{3},y,\beta_{3}}\left(z_{3},\bar{z}_{3}\right)\rangle.
\end{equation}

\section{The $n$-Graviton Amplitudes}

In this section, we generalise the above argument to encompass tree-level
MHV $n$-graviton amplitudes for $n\geq5$. We will demonstrate that
the corresponding celestial amplitude $\widehat{M}_{n}$ can be expressed
as the following correlation function of the family of operators $\mathcal{G}_{\varepsilon,y,\Delta}\left(z,\bar{z}\right)$
and $\mathcal{H}_{\varepsilon,y,\Delta}\left(z,\bar{z}\right)$, which
were introduced in Eqs. (\ref{eq:Operator-G}, \ref{eq:Operator-H}),
in the large-$N$ semiclassical $b\rightarrow0$ limit:
\begin{align}
\widehat{M}_{n} & =\delta_{\beta}\int\left[d\theta dyd\lambda\right]\delta_{y}\langle\mathcal{G}_{\varepsilon_{1},y,\Delta_{1}}\left(z_{1},\bar{z}_{1}\right)\prod_{k=2}^{n-2}\mathcal{H}_{\varepsilon_{k},y,\Delta_{k}}\left(z_{k},\bar{z}_{k}\right)\label{eq:n-Point-Celestial}\\
 & \mathcal{G}_{\varepsilon_{n-1},y,\Delta_{n-1}}\left(z_{n-1},\bar{z}_{n-1}\right)\mathcal{G}_{\varepsilon_{n},y,\Delta_{n}}\left(z_{n},\bar{z}_{n}\right)\rangle+\mathcal{P}\left(2,3,...,n\right).
\end{align}
Here, $\mathcal{P}\left(2,3,...,n\right)$ signifies permutation with
respect to the indices belonging to the set $\{2,3,...,n\}$, and
$\delta_{\beta}$ represents the delta function of the conformal weights
$\beta_{k}=\beta_{k}\left(\Delta_{k}\right)$, as defined in Eq. (\ref{eq:Delta-Conformal}).

As in the preceding section, let $u_{1}^{A}$, $u_{2}^{A}$, ... and
$\bar{u}_{1}^{\dot{A}}$, $\bar{u}_{2}^{\dot{A}}$, ... denote two
sequences of two-component spinors, while $p_{1}^{\mu}$, $p_{2}^{\mu}$,
... represent a family of null four-vectors corresponding to the momenta
of the gravitons. These are related by $p_{i}=\omega_{i}q_{i}$, where
$\omega_{i}$ denotes the frequency of the $i^{\text{th}}$ graviton,
and $q_{i}^{\mu}=(\sigma^{\mu})_{A\dot{A}}u_{i}^{A}\bar{u}_{i}^{B}$.
Using the little group rescaling property, we can fix the parametrisation
$u_{i}^{A}=\sqrt[]{\omega_{i}}\left(z_{i},1\right)^{T}$ and $\bar{u}_{i}^{\dot{A}}=\sqrt[]{\omega_{i}}\left(\bar{z}_{i},1\right)^{T}$,
where $z_{i}$ and $\bar{z}$$_{i}$ are local coordinates on $\mathbf{CP}^{1}$.
In Appendix B, we showed that the tree-level MHV $n$-graviton amplitude
$M_{n}$ (for $n\geq5$) can be expressed as:\emph{
\begin{equation}
M_{n}=-\frac{\omega_{1}^{3}\omega_{2}^{3}}{\omega_{3}\omega_{4}...\omega_{n}}\frac{z_{12}^{8}}{z_{12}z_{23}...z_{n1}}\int\left[dxd\lambda\right]\langle\lambda\big|\mathcal{Q}_{1}\prod_{k=2}^{n-2}\mathcal{P}_{k}\mathcal{Q}_{n-1}\mathcal{Q}_{n}\bar{\partial}\big|\lambda\rangle+\mathcal{P}\left(2,3,...,n-2\right),\label{eq:n-Graviton}
\end{equation}
}where $\int\left[dxd\lambda\right]\coloneqq\int\frac{d^{4}x}{\left(2\pi\right)^{4}}\int\frac{d^{2}z_{\lambda}}{\pi}$,
$C\left(n\right)\coloneqq\left\langle 12\right\rangle \left\langle 23\right\rangle ...\left\langle \left(n-1\right)n\right\rangle $
is the cyclic permutation, $\big|\lambda\rangle\coloneqq\chi^{\dagger}\left(\lambda\right)\big|0\rangle$
and $\bar{\partial}\coloneqq\partial/\partial\bar{\lambda}$. Here,
$\lambda^{A}\in\mathbf{C}^{2}$ is an auxiliary two-component spinor.
The fermionic doublets $(\chi\left(u\right)$, $\chi^{\dagger}$$\left(u\right))$
and $(\hat{\chi}\left(z\right)$, $\hat{\chi}^{\dagger}\left(z\right))$
were introduced in Eq. (\ref{eq:Mode}), and the operators $\mathcal{Q}_{i}\coloneqq\mathcal{Q}\left(u_{i},\bar{u}_{i}\right)$
and $\mathcal{P}_{i}\coloneqq\mathcal{P}\left(u_{i},\bar{u}_{i}\right)$
were defined in Eq. (\ref{eq:Q-P-Operators}). 

The $\epsilon$-regulated Mellin transform of Eq. (\ref{eq:n-Graviton})
is the celestial amplitude associated to $M_{n}$, and can be written
as:
\begin{equation}
\widehat{M}_{n}=-\frac{z_{12}^{8}}{z_{12}z_{23}...z_{n1}}\int\left[dxd\lambda\right]\langle\lambda\big|\widehat{\mathcal{Q}}_{1}\prod_{k=2}^{n-2}\widehat{\mathcal{P}}_{k}\widehat{\mathcal{Q}}_{n-2}\widehat{\mathcal{Q}}_{n-1}\bar{\partial}_{\lambda}\big|\lambda\rangle+\mathcal{P}\left(2,3,...,n\right),\label{eq:Celestial}
\end{equation}
where $\widehat{\mathcal{Q}}_{k}=\widehat{\mathcal{Q}}_{\beta_{k}}\left(u_{k},\bar{u}_{k}\right)$
and $\widehat{\mathcal{P}}_{k}=\widehat{\mathcal{P}}_{\beta_{k}}\left(u_{k},\bar{u}_{k}\right)$
are the $\epsilon$-regulated Mellin transformed operators as calculated
in Eqs. (\ref{eq:Mellin-transformed-operator-1}, \ref{eq:Mellin-transformed-operator-2}),
with conformal weights:
\begin{equation}
\beta_{k}=\Delta_{k}+\alpha_{k}-1\,\left(k=1,n-1,n\right),\,\,\,\beta_{k}=\Delta_{k}+\alpha_{k}\,\left(k=2,...,n-2\right).
\end{equation}
Here, $\alpha_{1}=\alpha_{n-1}=\alpha_{n}=1$ and $\alpha_{i}=0$
(for $i=2,3,...,n-2$), representing the powers of the frequencies
$\omega_{k}$ $\left(k=1,2,...\right)$ that factor into the first
term of Eq. (\ref{eq:n-Graviton}).

The next phase of our argument is based upon the results found in
Appendix C, where we established the following identity:
\begin{equation}
\mathbf{X}_{1}\mathbf{X}_{2}...\mathbf{X}_{n}G_{n+1}G_{n+2}=\int dy\,\delta_{y}\mathcal{D}_{1}\mathcal{D}_{2}...\mathcal{D}_{n}\mathcal{T}_{n+1}\mathcal{T}_{n+2}\prod_{k=1}^{n+2}K_{k,\beta_{k}}\left(x\right).\label{eq:Middle-argument}
\end{equation}
Here, $\mathbf{X}_{i}\left(x\right)$, $\mathcal{D}_{i}\left(y\right)$
and $\mathcal{T}_{i}\left(y\right)$ are the operators defined in
Eqs. (\ref{eq:Operator-X}, \ref{eq:Operators-DT}). The functions
$G_{\beta_{i}}\left(x,q_{i}\right)$ and $K_{i,\beta_{i}}\left(x,q\right)$
were introduced in Eq. (\ref{eq:Functions}). For the sake of simplicity,
we denote $\int dy\delta_{y}\coloneqq\int d^{4}y\delta^{\left(4\right)}\left(y\right)$,
and henceforth omit the variable $y$ from the argument.

Now, observe that $\widehat{\mathcal{Q}}_{i}=G_{i}\hat{\chi}^{\dagger}\left(z_{i}\right)\hat{\chi}\left(z_{i}\right)$
and $\widehat{\mathcal{P}}_{i}=\mathbf{X}_{i}$ for all $i=1,2,...,n$,
under the condition that we choose $M_{i}^{A\dot{A}}=-i\lambda^{A}\bar{u_{i}}/\left\langle u\lambda\right\rangle $.
By defining $\mathcal{R}_{i,y}\coloneqq\mathcal{T}_{i}\left(y\right)$
$\hat{\chi}^{\dagger}\left(z_{i}\right)\hat{\chi}\left(z_{i}\right)$,
Eq. (\ref{eq:Middle-argument}) yields:
\begin{equation}
\prod_{k=2}^{n-2}\widehat{\mathcal{P}}_{k}\widehat{\mathcal{Q}}_{n-1}\widehat{\mathcal{Q}}_{n}=\int dy\delta_{y}\prod_{i=1}^{n-2}\mathcal{D}_{i}\mathcal{R}_{n-1}\mathcal{R}_{n}\prod_{k=2}^{n}K_{k,\beta_{k}}\left(x\right).
\end{equation}
Thus, Eq. (\ref{eq:Celestial}) becomes:
\begin{align}
\widehat{M}_{n} & =-\frac{z_{12}^{8}}{z_{12}z_{23}...z_{n1}}\int\left[dydz_{\lambda}\right]\delta_{y}\langle\lambda\big|\mathcal{D}_{1}\prod_{i=1}^{n-2}\mathcal{D}_{i}\mathcal{R}_{n-1}\mathcal{R}_{n}\bar{\partial}_{\lambda}\big|\lambda\rangle\int\left[dx\right]\prod_{k=1}^{n}\frac{\Gamma\left(\beta_{k}\right)}{\left(\varepsilon_{k}-iq_{k}x\right)^{\beta_{k}}}\label{eq:Celestial-1}\\
 & +\mathcal{P}\left(2,3,...,n\right).
\end{align}

To evaluate the integral:
\begin{equation}
\int\left[dx\right]\prod_{k=1}^{n}\frac{\Gamma\left(\beta_{k}\right)}{\left(\varepsilon_{k}-iq_{k}x\right)^{\beta_{k}}},\label{eq:Integral}
\end{equation}
we begin by introducing an infinitesimal parameter $\epsilon$, replacing
$\varepsilon_{k}$ with $\epsilon\varepsilon_{k}$ in the integrand,
and then taking the limit as $\epsilon\rightarrow0$, while treating
$\varepsilon_{k}$ $\left(k\geq1\right)$ as independent variables.
This substitution transforms Eq. (\ref{eq:Integral}) into:
\begin{equation}
\int\left[dx\right]\prod_{k=1}^{n}\frac{\Gamma\left(\beta_{k}\right)}{\left(\varepsilon_{k}-iq_{k}x\right)^{\beta_{k}}}=\int\left[dx\right]\prod_{k=1}^{n}\frac{e^{-\beta_{k}\log\varepsilon_{k}}\Gamma\left(\beta_{k}\right)}{\left(\epsilon-i\ell_{k}x\right)^{\beta_{k}}}.
\end{equation}
As in the preceding section, we define $\ell_{k}^{\mu}$ to be the
four-vector given by $\ell_{k}^{\mu}\coloneqq(\sigma^{\mu})_{A\dot{A}}\pi_{k}^{A}\bar{\pi}_{k}^{\dot{A}}$,
where $\pi_{k}^{A}$ and $\bar{\pi}_{k}^{\dot{A}}$ are two-component
spinors parametrised in the local coordinates $z_{k},\bar{z}_{k}$
$\in\mathbf{CP}^{1}$ by $\pi_{k}^{A}=\varepsilon_{k}^{-1/2}\left(z_{k},1\right)^{T}$
and $\bar{\pi}_{k}^{\dot{A}}=\varepsilon_{k}^{-1/2}\left(\bar{z}_{k},1\right)^{T}$,
using the little group reparametrization property. Applying the results
from Appendix A, we obtain, in the semiclassical $b\rightarrow0$
limit of Liouville theory:
\begin{equation}
\int\left[dx\right]\prod_{k=1}^{n}\frac{\Gamma\left(\beta_{k}\right)}{\left(\varepsilon_{k}-iq_{k}x\right)^{\beta_{k}}}=\delta_{\beta}\left\langle \prod_{k=1}^{n}e^{-\beta_{k}\log\varepsilon_{j}}\Gamma\left(\beta_{k}\right)V_{b\beta_{k}}\left(\frac{z_{k}}{\sqrt{\varepsilon_{k}}},\frac{\bar{z}_{k}}{\sqrt{\varepsilon_{k}}}\right)\right\rangle +\left(\bar{z}\leftrightarrow-\bar{z}\right),\label{eq:Argument-2-2}
\end{equation}
where the normalised delta function of the conformal weights is:
\begin{equation}
\delta_{\beta}\coloneqq\frac{1}{8\pi^{3}\mathcal{N}_{\mu,b}}\delta\left(4-\sum_{r=1}^{n}\beta_{r}\right),
\end{equation}
and $V_{\alpha}\left(z,\bar{z}\right)=:e^{2\alpha\hat{\phi}\left(z,\bar{z}\right)}:$
is the Liouville vertex operator with momentum $\alpha$. As explained
earlier, the symbol $\left(\bar{z}\leftrightarrow-\bar{z}\right)$
indicates that the first term is replicated with the substitution
of $\bar{z}$ by $-\bar{z}$.

Consequently, Eq. (\ref{eq:Celestial-1}) is equivalent to:
\begin{align}
\widehat{M}_{n} & =\delta_{\beta}\frac{z_{12}^{8}}{z_{12}z_{23}...z_{n1}}\int\left[d\theta dydz_{\lambda}\right]\delta_{y}\langle\lambda\big|\mathcal{D}_{1}\prod_{i=1}^{n-2}\mathcal{D}_{i}\mathcal{R}_{n-1}\mathcal{R}_{n}\bar{\partial}_{\lambda}\big|\lambda\rangle\\
 & \left\langle \prod_{k=1}^{n}e^{-\beta_{k}\log\varepsilon_{k}}\Gamma\left(\beta_{k}\right)V_{b\beta_{k}}\left(\frac{z_{k}}{\sqrt{\varepsilon_{k}}},\frac{\bar{z}_{k}}{\sqrt{\varepsilon_{k}}}\right)\right\rangle +\mathcal{P}\left(2,3,...,n\right)+\left(\bar{z}\leftrightarrow-\bar{z}\right),\label{eq:Celestial-quasi-final}
\end{align}
in the limit $b\rightarrow0$.

Finally, let us consider the matter system consisting of chiral fermionic
primary free fields $\psi^{i}$ $\left(i=1,2,...,N\right)$ on $\mathbf{CP}^{1}$
transforming under the adjoint representation of the group $SO\left(N\right)$--obeying
the OPEs defined in Eq. (\ref{eq:OPE})--and let $J^{a}\left(z\right)$
be the level-one $SO\left(N\right)$ Kac-Moody currents defined in
Eq. (\ref{eq:Currents}). Then, Eq. (\ref{eq:Celestial-quasi-final})
can be written as:
\begin{align}
\widehat{M}_{n} & =-\delta_{\beta}\int\left[dydz_{\lambda}\right]\int\left[d\theta dyd\lambda\right]\delta_{y}\left\langle \prod_{r=1}^{n}J\left(z_{r}\right)\right\rangle \\
 & \langle\lambda\big|\mathcal{D}_{1}\prod_{i=1}^{n-2}\mathcal{D}_{i}\mathcal{R}_{n-1}\mathcal{R}_{n}\bar{\partial}_{\lambda}\big|\lambda\rangle\left\langle \prod_{k=1}^{n}e^{-\beta_{k}\log\varepsilon_{k}}\Gamma\left(\beta_{k}\right)V_{b\beta_{k}}\left(\frac{z_{k}}{\sqrt{\varepsilon_{k}}},\frac{\bar{z}_{k}}{\sqrt{\varepsilon_{k}}}\right)\right\rangle \\
 & +\mathcal{P}\left(2,3,...,n\right)+\left(\bar{z}\leftrightarrow-\bar{z}\right),
\end{align}
in the large-$N$ semiclassical $b\rightarrow0$ limit, which is clearly
equivalent to Eq. (\ref{eq:n-Point-Celestial}) upon the substitution
of the definitions of the operators $\mathcal{G}_{\varepsilon,y,\beta}\left(z,\bar{z}\right)$
and $\mathcal{H}_{\varepsilon,y,\beta}\left(z,\bar{z}\right)$ defined
in Eqs. (\ref{eq:Operator-G}, \ref{eq:Operator-H}).

\section{Discussion}

In this note, we have demonstrated that tree-level MHV $n$-graviton
amplitudes in flat space, upon analytical continuation to Klein's
signature, can be holographically generated by a particular limit
of a $2d$ CFT. This $2d$ CFT comprises a Liouville field theory
with a central charge $c=1+6Q^{2}$, where $Q=b+b^{-1}$, a level-one
$SO\left(N\right)$ Kac-Moody current, and a doublet of free fermions
on $\mathbf{CP}^{1}$ (in addition to a pair of spinor fields incorporated
through a Berezin integral). The limit in which this $2d$ CFT is
shown to generate $n$-graviton amplitudes corresponds to $N\rightarrow\infty$
and $b\rightarrow0$. It is essential to note that this system, while
not unique or particularly elegant, serves primarily as a toy model
to explore the celestial holography dictionary. We emphasise that
the intent of this work is not to establish a comprehensive holographic
duality between a $2d$ CFT and four-dimensional quantum gravity in
flat spacetime. Rather, it seeks to define a correspondence between
specific \emph{limiting subsector} of a $2d$ CFT and analogous subsector
within four-dimensional gravity in flat space.

The findings presented herein open several avenues for further exploration,
and we hope that these will inspire subsequent research endeavours
by ourselves and others in the community. Some of the open problems
that we are currently investigating are as follows:
\begin{enumerate}
\item Our $2d$ CFT is presently capable of reproducing only MHV amplitudes.
In \citet{hodges2012simple}, expressions for non-MHV (NMHV) amplitudes
using twistor variables were examined. It would be intriguing to explore
whether our CFT model, or a suitable extension thereof, could successfully
reproduce these NMHV amplitudes.
\item A potential critique of our construction is that our CFT correlators,
particularly the freedom in choosing celestial graviton operators,
merely reorganise the variables in which the $n$-point amplitudes
are expressed. In our view, the most effective response to this critique
would be to provide a dynamical argument that justifies the connection
between our $2d$ CFT and Einstein's gravity in flat space.In \citet{stieberger2023yang},
the authors investigated the asymptotic behaviour of Liouville correlators
beyond the semiclassical limit, utilising the asymptotic properties
of the DOZZ formula, as studied, for example, in \citet{harlow2011analytic}.
Notably, \citet{stieberger2023celestial} proposed a relation between
the Liouville coupling constant $b$, the Yang-Mills coupling $g\left(\Lambda\right)$
at renormalisation scale $\Lambda$, and the one-loop coefficient
$\beta_{0}$ of the Yang-Mills beta function:
\begin{equation}
b^{2}=\frac{\beta_{0}}{8\pi^{2}}g^{2}\left(M\right).
\end{equation}
It would be highly interesting to investigate whether ow expression
(Eq. (\ref{eq:n-Point-Celestial})) for the celestial $n$-graviton
amplitude $\widehat{M}_{n}$ in terms of our $2d$ CFT correlators
could be employed to derive an analogous formula for gravity, possibly
drawing from ideas such as those in the asymptotic safety program
(\citet{niedermaier2006asymptotic}).
\item The OPEs of the celestial graviton operators defined in Eqs. are determined
entirely by the corresponding OPEs of the level-one $SO\left(N\right)$
Kac-Moody currents (Eq. (\ref{eq:OPE})) and the mode expansion of
the fermionic doublet $\chi$, $\chi^{\dagger}$ (Eq. (\ref{eq:Mode})).
However, it has been shown in the works of \citet{pate2019conformally,guevara2021holographic,himwich2022celestial,strominger2021w,banerjee2021mhv}
that asymptotic symmetries, or the soft collinear limit of celestial
amplitudes, dictate the celestial OPEs. Investigating the connection
between the OPEs obeyed by our celestial graviton operators and the
algebraic structures uncovered by these authors would be of considerable
interest.
\item An additional line of inquiry involves extending the present analysis
from Einstein's gravity to Einstein-Yang-Mills theory in flat space.
\end{enumerate}

\section{Acknowledgement}

This research was supported in part by Perimeter Institute for Theoretical
Physics. Research at Perimeter Institute is supported by the Government
of Canada through the Department of Innovation, Science and Economic
Development and by the Province of Ontario through the Ministry of
Research, Innovation and Science. The author also thanks Giorgio Torrieri
for his important encouragement.

\appendix

\section{Frequently Used Integrals in Mellin and Klein Spaces\label{sec:Frequently-used-integrals}}

In the computation of $n$--point celestial amplitudes, the following
integral in Klein space within the Mellin representation frequently
appears:
\begin{equation}
F_{n}\left(\left\{ \Delta_{i},z_{i},\bar{z}_{i}\right\} \big|\left\{ \alpha_{j}\right\} \right)=\int\frac{d^{4}X}{\left(2\pi\right)^{4}}\left(\prod_{k=1}^{n}\int_{0}^{\infty}d\omega_{k}\,\,\,\omega_{k}^{\Delta_{k}-1}e^{-\varepsilon\omega_{k}}e^{ip_{k}\cdot X}\right)\frac{1}{\omega^{\alpha_{1}}\omega^{\alpha_{2}}...\omega^{\alpha_{n}}}.\label{eq:MKI-1}
\end{equation}
Here, $\varepsilon>0$ is an infinitesimal regulator, and $\alpha_{i}\geq1$
($i=1,...,n$) is an integer. The indices $i$, $j$ and $k$ run
from $1$ to $n$. The notation $\left\{ x_{i},y_{i},z_{i},...\right\} $
represents the collection of all variables $x_{1},y_{1},z_{1}$, ...,
$x_{n},y_{n},z_{n}$.

To begin our computation, recall that the standard null-momentum in
Klein signature is defined as:
\begin{equation}
\hat{p}_{k}^{\mu}\left(z_{k},\bar{z}_{k}\right)=\left(1-z_{k}\bar{z}_{k},z_{k}+\bar{z}_{k},1+z_{k}\bar{z}_{k},z_{k}-\bar{z}_{k}\right).
\end{equation}
So, any light-like momentum can be expressed as $p_{k}=\varepsilon_{k}\omega_{k}\hat{p}_{k}$.
Here, $\omega_{k}>0$ is the particle's frequency, and $\varepsilon_{k}=-1$
for incoming particles and $\varepsilon_{k}=1$ for incoming particles.
With this, Eq. (\ref{eq:MKI-1}) can be rewritten as:
\begin{equation}
F_{n}=\int\frac{d^{4}X}{\left(2\pi\right)^{4}}\prod_{k=1}^{n}\int_{0}^{\infty}d\omega_{k}\,\,\,\omega^{\left(\Delta_{k}-\alpha_{k}\right)-1}e^{-\omega_{k}\left(\varepsilon-i\hat{p}_{k}\cdot X\right)}.
\end{equation}

The next step of the derivation is to define new variables $t_{k}=\omega_{k}\left(\varepsilon-i\hat{p}_{k}\cdot X\right)$
and apply this change of variables to the integral:
\begin{align}
F_{n} & =\int\frac{d^{4}X}{\left(2\pi\right)^{4}}\prod_{k=1}^{n}\frac{1}{\left(\epsilon-i\varepsilon_{k}\hat{p}_{k}\left(z_{k},\bar{z}_{k}\right)\cdot X\right)^{\Delta_{k}-\alpha_{k}}}\int_{0}^{\infty}dt_{k}\,\,\,t_{k}^{\left(\Delta_{k}-\alpha_{k}\right)-1}e^{-t_{k}}\\
 & =\frac{\prod_{i=1}^{n}\Gamma\left(\Delta_{i}-\alpha_{i}\right)}{\left(2\pi\right)^{4}}\int d^{4}X\,\,\,\prod_{k=1}^{n}\frac{1}{\left(\epsilon-i\varepsilon_{k}\hat{p}_{k}\left(z_{k},\bar{z}_{k}\right)\cdot X\right)^{\Delta_{k}-\alpha_{k}}}.\label{eq:MKI-2}
\end{align}

Following\footnote{See \citet{cheung20174d} for an alternative spacetime foliation.}
\citet{melton2023celestial}, the time-like wedge of Klein space (where
$X^{2}<0$) is parametrised by $X=\tau\hat{x}_{+}$, while the space-like
wedge (where $X^{2}>0$) is parametrised by $X=\tau\hat{x}_{-}$.
Here, $\hat{x}_{+}^{2}=-1$ and $\hat{x}_{-}^{2}=1$, with $\tau>0$
representing the ``length'' of a world-vector in Klein signature \citet{barrett1994kleinian}.
Therefore, Eq. (\ref{eq:MKI-2}) becomes:
\begin{align}
 & F_{n}=\frac{\prod_{i=1}^{n}\Gamma\left(\Delta_{i}-\alpha_{i}\right)}{\left(2\pi\right)^{4}}\int_{0}^{\infty}d\tau\,\,\,\tau^{3}\Bigg[\int_{\hat{x}_{-}^{2}=1}d^{3}\hat{x}_{-}\prod_{j=1}^{n}\frac{1}{\left(\epsilon-i\varepsilon_{k}\tau\hat{p}_{k}\left(z_{k},\bar{z}_{k}\right)\cdot\hat{x}_{-}\right)^{\Delta_{k}-\alpha_{k}}}\\
 & \int_{\hat{x}_{+}^{2}=-1}d^{3}\hat{x}_{+}\prod_{j=1}^{n}\frac{1}{\left(\epsilon-i\varepsilon_{k}\tau\hat{p}_{k}\left(z_{k},\bar{z}_{k}\right)\cdot\hat{x}_{+}\right)^{\Delta_{k}-\alpha_{k}}}\Bigg].
\end{align}

Note that the second integral inside the brackets can be obtained
from the first by replacing $\bar{z}_{k}\mapsto-\bar{z}_{k}$. Thus,
$F_{n}$ can be expressed like a sum ``reflected amplitudes'' with
respect to $\bar{z}_{k}$:
\begin{equation}
F_{n}=G_{n}\left(\left\{ \Delta_{i},z_{i},\bar{z}_{i}\right\} \big|\left\{ \alpha_{j}\right\} \right)+G_{n}\left(\left\{ \Delta_{i},z_{i},-\bar{z}_{i}\right\} \big|\left\{ \alpha_{j}\right\} \right),\label{eq:MKI-3}
\end{equation}
where the new quantity $G_{n}$ is defined as:
\begin{equation}
G_{n}=\frac{\prod_{i=1}^{n}\Gamma\left(\Delta_{i}-\alpha_{i}\right)}{\left(2\pi\right)^{4}}\int_{0}^{\infty}d\tau\,\,\,\tau^{3}\int_{AdS_{3}/\mathbb{Z}}d^{3}\hat{x}\prod_{j=1}^{n}\frac{1}{\left(\epsilon-i\varepsilon_{k}\tau\hat{p}_{k}\left(z_{k},\bar{z}_{k}\right)\cdot\hat{x}\right)^{\Delta_{k}-\alpha_{k}}}.\label{eq:MKI-4}
\end{equation}
Here, $d^{3}\hat{x}$ is the measure on Lorentzian $AdS_{3}$ with
periodic time.

Recall that $\epsilon$ is a positive infinitesimal introduced as
a regulator of Mellin integral. If $\epsilon'=\epsilon/\tau$, then
$\epsilon'$ is also a positive infinitesimal. Therefore, we can express
Eq. (\ref{eq:MKI-4}) as:
\begin{align}
G_{n} & =\frac{\prod_{i=1}^{n}\Gamma\left(\Delta_{i}-\alpha_{i}\right)}{\left(2\pi\right)^{4}}\int_{AdS_{3}/\mathbb{Z}}d^{3}\hat{x}\prod_{j=1}^{n}\frac{1}{\left(\epsilon'-i\varepsilon_{k}\hat{p}_{k}\left(z_{k},\bar{z}_{k}\right)\cdot\hat{x}\right)^{\Delta_{k}-\alpha_{k}}}\int_{0}^{\infty}d\tau\,\,\,\tau^{\left(4-\sum_{j}\left(\Delta_{j}-\alpha_{j}\right)\right)+1}\\
= & \frac{\prod_{i=1}^{n}\Gamma\left(\Delta_{i}-\alpha_{i}\right)}{8\pi^{3}}\delta\left(4-\sum_{j=1}^{n}\left(\Delta_{j}-\alpha_{j}\right)\right)\int_{AdS_{3}/\mathbb{Z}}d^{3}\hat{x}\prod_{j=1}^{n}\frac{1}{\left(\epsilon'-i\varepsilon_{k}\hat{p}_{k}\left(z_{k},\bar{z}_{k}\right)\cdot\hat{x}\right)^{\Delta_{k}-\alpha_{k}}},\label{eq:MKI-5}
\end{align}
since:
\begin{equation}
\int_{0}^{\infty}d\tau\,\,\,\tau^{-\beta-1}=2\pi\delta\left(\beta\right).
\end{equation}

Similarly, noting that $\varepsilon_{k}=\exp\left(-i\pi\left(1-\varepsilon_{k}\right)/2\right)$,
we can simplify Eq. (\ref{eq:MKI-5}) further as:
\begin{align}
G_{n} & =\frac{1}{8\pi^{3}}\prod_{i=1}^{n}\left(e^{-\frac{i\pi}{2}\left(1-\varepsilon_{i}\right)\left(\Delta_{i}-\alpha_{i}\right)}\Gamma\left(\Delta_{i}-\alpha_{i}\right)\right)\\
 & \times\delta\left(4-\sum_{j=1}^{n}\left(\Delta_{j}-\alpha_{j}\right)\right)\int_{AdS_{3}/\mathbb{Z}}d^{3}\hat{x}\prod_{j=1}^{n}\frac{1}{\left(\epsilon'-i\hat{p}_{k}\left(z_{k},\bar{z}_{k}\right)\cdot\hat{x}\right)^{\Delta_{k}-\alpha_{k}}}.
\end{align}

Finally, Melton \emph{et al. }\citet[App. A]{melton2024celestial}
proved that the integral above, which represents a contact Witten
diagram for massless particles \citet{penedones2017tasi}, can be
expressed as the semi-classical limit of the correlation function
of Liouville vertex operators $\mathcal{V}_{\alpha}\left(z,\bar{z}\right)=\mathbf{N}\big(e^{2\alpha\hat{\phi}\left(z,\bar{z}\right)}\big)$,
\begin{equation}
\int_{AdS_{3}/\mathbb{Z}}d^{3}\hat{x}\prod_{j=1}^{n}\frac{1}{\left(\epsilon'-i\hat{p}_{k}\left(z_{k},\bar{z}_{k}\right)\cdot\hat{x}\right)^{\Delta_{k}-\alpha_{k}}}=\frac{1}{\mathcal{N}_{\mu,b}}\left\langle \prod_{k=1}^{n}\mathcal{V}_{b\left(\Delta_{k}-1\right)}\left(z_{k},\bar{z}_{k}\right)\right\rangle .
\end{equation}
This result involves the constant: 
\begin{equation}
\mathcal{N}_{\mu,b}\coloneqq\frac{1}{\pi b^{3}}\exp\left(-2\gamma_{E}+\frac{2}{b^{2}}+\left(\frac{1}{b^{2}}-1-\frac{\beta}{2}\right)\log\lambda\right)\csc\left[\pi\left(\frac{1}{b^{2}}-\frac{\beta}{2}\right)\right],\label{eq:Constant}
\end{equation}
that depends on the Liouville ``cosmological constant'' $\mu$ and
the parameter $b$, which controls the semi-classical limit. 

Therefore:
\begin{equation}
G_{n}=\frac{1}{8\pi^{3}\mathcal{N}_{\mu,b}}\delta\left(4-\sum_{i=1}^{n}\left(\Delta_{i}-\alpha_{i}\right)\right)\left\langle \prod_{j=1}^{n}e^{-\frac{i\pi}{2}\left(1-\varepsilon_{j}\right)\left(\Delta_{j}-\alpha_{j}\right)}\Gamma\left(\Delta_{j}-\alpha_{j}\right)\mathcal{V}_{b\left(\Delta_{j}-1\right)}\left(z_{j},\bar{z}_{j}\right)\right\rangle .
\end{equation}

\section{The $n$-Point Graviton Amplitude Expressed as Correlators of $\mathbf{CP}^{1}$
Fields}

In this Appendix, we will demonstrate a key mathematical result employed
in Section 3, which enabled the extension of the argument presented
in Section 2---wherein the tree-level MHV four-graviton amplitude
was expressed as correlators of a $2d$ CFT---to encompass the general
case of $n$-graviton amplitudes: 

\textbf{Proposition}. Let $M_{n}$ represent the tree-level MHV $n$-graviton
amplitude, and let $\hat{\chi}$ and $\hat{\chi}^{\dagger}$ be a
doublet of fermionic fields defined on $\mathbf{CP}^{1}$, with mode
expansions given by:
\begin{equation}
\hat{\chi}\left(z\right)\coloneqq\sum_{n=0}^{n}b_{n}z^{n},\,\,\,\hat{\chi}^{\dagger}\left(z\right)\coloneqq\sum_{n=0}^{\infty}b_{n}^{\dagger}z^{-n-1},\label{eq:Modes-2}
\end{equation}
where $b_{n}$ and $b_{n}^{\dagger}$ are fermionic creation and annihilation
operators, satisfying the anti-commutation relations $\{b_{m},b_{n}^{\dagger}\}=\delta_{mn}$.
The vacuum state $|0\rangle$ is defined by the condition $b_{m}|0\rangle=0$
for all $m\geq0$. Let $u^{A}=\left(\xi,\zeta\right)^{T}\in\mathbf{C}^{2}$
be a two-component spinor such that $\zeta\neq0$, and $z\coloneqq\xi/\zeta\in\mathbf{CP}^{1}$
be its corresponding local coordinate on the complex projective line.
Then, we define the pair of fermions $\chi$ and $\chi^{\dagger}$
on the space of two-component spinors as follows:
\begin{equation}
\chi\left(u\right)\coloneqq\zeta^{-1}\hat{\chi}\left(u\right),\,\,\,\chi^{\dagger}\left(u\right)\coloneqq\zeta^{-1}\hat{\chi}^{\dagger}\left(z\right).
\end{equation}

Let $x$ be a four-vector in spacetime, and $p$, $p_{1}$, $p_{2}$,
... be a sequence of light-like four-vectors representing the momenta
of the gravitons with frequencies $\omega$, $\omega_{1}$, $\omega_{2}$,
..., such that $p=\omega q$ and $p_{k}=\omega_{k}q_{k}$ for all
$k\geq1$, where $q$ and $q_{k}$ are the standard null momenta,
which we parametrise (using, if necessary, the little group rescaling)
by $q^{\mu}=(\sigma^{\mu})_{A\dot{A}}v^{A}\bar{v}^{\dot{A}}$ and
$q_{k}^{\mu}=(\sigma^{\mu})_{A\dot{A}}v_{k}^{A}\bar{v}_{k}^{\dot{A}}$.
For convenience, we choose $v_{k}^{\dot{A}}\coloneqq\left(z,1\right)^{T}$
and $\bar{v}_{k}^{\dot{A}}\coloneqq\left(\bar{z},1\right)$. Moreover,
let $u$, $u_{1}$, $u_{2}$, ... and $\bar{u}$, $\bar{u}_{1}$,
$\bar{u}_{2}$, ... be two families of two-component spinors such
that $p^{\mu}=(\sigma^{\mu})_{A\dot{A}}u^{A}\bar{u}^{\dot{A}}$ and
$p_{k}^{\mu}=(\sigma^{\mu})_{A\dot{A}}u_{k}^{A}\bar{u}_{k}^{\dot{A}}$
for all $k\geq1$\emph{, }and define the operators $\mathcal{Q}\left(u,\bar{u}\right)$
and $\mathcal{P}\left(u,\bar{u}\right)$ by:\emph{
\begin{equation}
\mathcal{Q}\left(u,\bar{u}\right)\coloneqq e^{ip\cdot x}\chi^{\dagger}\left(u\right)\chi\left(u\right),\,\,\,\mathcal{P}\left(u,\bar{u}\right)\coloneqq-i\frac{\lambda^{A}\bar{u}^{\dot{A}}}{\left\langle u\lambda\right\rangle }\frac{\partial}{\partial x^{A\dot{A}}}e^{ip\cdot x},\label{eq:Operators}
\end{equation}
}where $\lambda^{A}$ is a fixed spinor. 

Therefore, by defining the state $\big|\lambda\rangle\coloneqq\chi^{\dagger}\left(\lambda\right)\big|0\rangle$,
the tree-level MHV scattering amplitudes for $n=3$, $n=4$ and $n\geq5$
gravitons can be expressed respectively in terms of an integral representation
involving $\mathbf{CP}^{1}$ correlators:
\begin{equation}
M_{3}=-\frac{\left\langle 12\right\rangle ^{8}}{C\left(3\right)}\int\left[dxd\lambda\right]\,\langle\lambda\big|\mathcal{Q}_{1}\mathcal{Q}_{2}\mathcal{Q}_{3}\bar{\partial}\big|\lambda\rangle,
\end{equation}
\emph{
\begin{equation}
M_{4}=-\frac{\left\langle 12\right\rangle ^{8}}{C\left(4\right)}\int\left[dxd\lambda\right]\langle\lambda\big|\mathcal{Q}_{1}\mathcal{P}_{2}\mathcal{Q}_{3}\mathcal{Q}_{4}\bar{\partial}\big|\lambda\rangle,\label{eq:M4-1}
\end{equation}
}and, for $n\geq5$,\emph{
\begin{equation}
M_{n}=-\frac{\left\langle 12\right\rangle ^{8}}{C\left(n\right)}\int\left[dxd\lambda\right]\langle\lambda\big|\mathcal{Q}_{1}\prod_{k=2}^{n-2}\mathcal{P}_{k}\mathcal{Q}_{n-1}\mathcal{Q}_{n}\bar{\partial}\big|\lambda\rangle+\mathcal{P}\left(2,3,...,n-2\right)
\end{equation}
}where $C\left(n\right)\coloneqq\left\langle 12\right\rangle \left\langle 23\right\rangle ...\left\langle \left(n-1\right)n\right\rangle $
is the cyclic permutation (recognised as the Parke-Taylor denominator),
$\mathcal{P}\left(2,3,...,n-2\right)$ represents the permutation
of the indices $\{2,3,...,n-2\}$, and to simplify our notation we
define $\bar{\partial}\coloneqq\partial/\partial\bar{z}_{\lambda}$
and:
\begin{equation}
\int\left[dxd\lambda\right]\coloneqq\int\frac{d^{4}x}{\left(2\pi\right)^{4}}\int\frac{d^{2}z_{\lambda}}{\pi},\mathcal{Q}_{k}\coloneqq\mathcal{Q}\left(u_{k},\bar{u}_{k}\right),\,\,\,\mathcal{P}_{k}\coloneqq\mathcal{P}\left(u_{k},\bar{u}_{k}\right).
\end{equation}

To streamline our forthcoming analysis, we assume without loss of
generality that all particles are outgoing. This simplification serves
to make the calculations less cumbersome. For notational brevity,
we shall henceforth omit $\bar{u}$ in the arguments of $\mathcal{P}$,
and simply denote it as $\mathcal{P}\left(u\right)$, implicitly understanding
it as $\mathcal{P}\left(u,\bar{u}\right)$.

\subsection{$3$-Graviton and $4$-Graviton Amplitudes\label{subsec:M3-M4-Amplitudes}}

It is well-established in modern quantum field theory texts (e.g.,
\citet{elvang2013scattering,badger2024scattering}) that the tree-level
MHV $3$-point function for the scattering of gravitons $1^{-}$,
$2^{-}$ and $3^{+}$ is given by:
\begin{equation}
M_{3}=-\frac{\left\langle 12\right\rangle ^{8}}{C\left(3\right)}\frac{1}{\left\langle 12\right\rangle \left\langle 23\right\rangle \left\langle 34\right\rangle }\delta_{P},\label{eq:M3}
\end{equation}
where, for brevity, we denote the delta function associated with momentum
conservation as:
\begin{equation}
\delta_{P}\coloneqq\delta^{\left(4\right)}\left(P\right),\,\,\,P\coloneqq p_{1}+p_{2}+p_{3}.
\end{equation}
Using the definition of the $\mathcal{Q}$ operator provided in Eq.
(\ref{eq:Operators}), it is a straightforward exercise to compute:
\begin{equation}
\langle\lambda\big|\mathcal{Q}_{1}\mathcal{Q}_{2}\mathcal{Q}_{3}\bar{\partial}\big|\lambda\rangle=e^{iP\cdot x}\frac{1}{\left\langle 12\right\rangle \left\langle 23\right\rangle \left\langle 31\right\rangle }\bar{\partial}\left(\frac{1}{\left\langle 3\lambda\right\rangle }\right).
\end{equation}
Integrating the above equation over the measure $\left[dxd\lambda\right]$,
and employing the identity $\bar{\partial}z^{-1}=\pi\delta\left(z\right)$,
we obtain:
\[
\int\left[dxd\lambda\right]\,\langle\lambda\big|\mathcal{Q}_{1}\mathcal{Q}_{2}\mathcal{Q}_{3}\bar{\partial}\big|\lambda\rangle=-\frac{1}{\left\langle 12\right\rangle \left\langle 23\right\rangle \left\langle 31\right\rangle }\delta_{P}.
\]
Thus, Eq. (\ref{eq:M3}) can be expressed as:
\begin{equation}
M_{3}=-\frac{\left\langle 12\right\rangle ^{8}}{C\left(3\right)}\int\left[dxd\lambda\right]\,\langle\lambda\big|\mathcal{Q}_{1}\mathcal{Q}_{2}\mathcal{Q}_{3}\bar{\partial}\big|\lambda\rangle.\label{eq:M3-final}
\end{equation}

According to \citet{berends1988relations}, the $4$-point function
for the scattering of four gravitons $1^{-}$, $2^{-}$, $3^{+}$
and $4^{+}$ is expressed as:
\begin{equation}
M_{4}=-\frac{\left\langle 12\right\rangle ^{8}}{C\left(4\right)}\frac{\left[12\right]}{\left\langle 34\right\rangle }\delta_{P},\label{eq:M4}
\end{equation}
where $P=p_{1}+p_{2}+p_{3}+p_{4}$. For simplicity, and without loss
of generality, let's assume that all gravitons are outgoing. Conservation
of momentum imposes the condition $[2\big|P\big|4\rangle=0$, which
implies that $\left[12\right]=\left[23\right]\left\langle 34\right\rangle /\left\langle 14\right\rangle $.
Additionally, $[2\big|p_{3}\big|4\rangle=\left[23\right]\left\langle 34\right\rangle $.
Consequently, Eq. (\ref{eq:M3}) can be rewritten as:
\begin{equation}
M_{4}=\frac{\left\langle 12\right\rangle ^{8}}{C\left(4\right)}\frac{1}{\left\langle 13\right\rangle \left\langle 14\right\rangle \left\langle 34\right\rangle }\frac{[2\big|p_{3}+p_{4}\big|4\rangle}{\left\langle 24\right\rangle }\delta_{P}.\label{eq:M4-Step}
\end{equation}
Since the $\mathcal{P}$ operator satisfies:
\begin{equation}
\mathcal{P}_{\ell}e^{ip_{\ell+1}}=e^{i\left(p_{\ell}+p_{\ell+1}\right)\cdot x}\frac{[\ell|p_{\ell+1}|\lambda\rangle}{\left\langle \ell\lambda\right\rangle },
\end{equation}
we find that:
\begin{equation}
\langle\lambda\big|\mathcal{Q}_{1}\mathcal{P}_{2}\mathcal{Q}_{3}\mathcal{Q}_{4}\bar{\partial}\big|\lambda\rangle=e^{iPx}\frac{1}{\left\langle \lambda1\right\rangle \left\langle 13\right\rangle \left\langle 34\right\rangle }\bar{\partial}\left(\frac{1}{\left\langle 4\lambda\right\rangle }\right)\frac{[2\big|p_{3}+p_{4}\big|\lambda\rangle}{\left\langle 2\lambda\right\rangle }.
\end{equation}
Integrating this expression with respect to the measure $\left[dxd\lambda\right]$,
employing again the identity $\bar{\partial}z^{-1}=\pi\delta\left(z\right)$,
and noting that $p_{4}$ is null (i.e., $\left\langle 44\right\rangle =0$),
we obtain:
\begin{equation}
\int\left[dxd\lambda\right]\,\langle\lambda\big|\mathcal{Q}_{1}\mathcal{P}_{2}\mathcal{Q}_{3}\mathcal{Q}_{4}\bar{\partial}\big|\lambda\rangle=-\frac{1}{\left\langle 13\right\rangle \left\langle 14\right\rangle \left\langle 34\right\rangle }\frac{[2\big|p_{3}+p_{4}\big|4\rangle}{\left\langle 24\right\rangle }\delta_{P}.
\end{equation}
From this result and Eq. (\ref{eq:M4-Step}), we conclude:
\begin{equation}
M_{4}=-\frac{\left\langle 12\right\rangle ^{8}}{C\left(4\right)}\int\left[dxd\lambda\right]\,\langle\lambda\big|\mathcal{Q}_{1}\mathcal{P}_{2}\mathcal{Q}_{3}\mathcal{Q}_{4}\bar{\partial}\big|\lambda\rangle.\label{eq:M4-final}
\end{equation}

\subsection{$n$-Graviton Amplitude for $n\protect\geq5$}

To develop an intuitive understanding of the correlators that appears
in the integral above, we first consider the initial examples in the
series of products of the form $\mathcal{P}_{\lambda}\mathcal{P}_{\lambda}...\mathcal{P}_{\lambda}e^{ip\cdot x}$.
Letting $p_{\ell}$, $p_{\ell+1}$, ... be null vectors, we find:
\begin{equation}
\mathcal{P}_{\lambda}\left(u_{\ell}\right)\mathcal{P}_{\lambda}\left(u_{\ell+1}\right)e^{ip_{\ell+2}\cdot x}=\frac{[\ell|p_{\ell+1}+p_{\ell+2}|\lambda\rangle}{\left\langle \ell\lambda\right\rangle }\frac{[\ell+1|p_{\ell+2}|\lambda\rangle}{\left\langle \left(\ell+1\right)\lambda\right\rangle }e^{i\left(p_{\ell}+p_{\ell+1}+p_{\ell+2}\right)\cdot x}.
\end{equation}

This leads us to the following induction hypothesis:
\begin{equation}
\prod_{k=\ell}^{m}\mathcal{P}_{\lambda}\left(u_{k}\right)e^{ip_{m+1}\cdot x}=e^{i\left(p_{\ell}+...+p_{m+1}\right)}\prod_{k=\ell}^{m}\frac{[k|p_{k+1}+...+p_{m+1}|\lambda\rangle}{\left\langle k\lambda\right\rangle }.\label{eq:Induction-Hypothesis}
\end{equation}

Assume the above equation holds for some $m=M\geq1$. We must now
demonstrate its validity for $m=M+1$. Indeed, we have:
\begin{align}
 & \prod_{k=\ell}^{M+1}\mathcal{P}_{\lambda}\left(u_{k}\right)e^{ip_{M+2}\cdot x}\\
 & =\mathcal{P}_{\lambda}\left(u_{\ell}\right)\left[\mathcal{P}_{\lambda}\left(u_{\ell+1}\right)...\mathcal{P}_{\lambda}\left(u_{M}\right)\mathcal{P}\left(u_{M+1}\right)e^{ip_{M+2}\cdot x}\right]\\
 & =\mathcal{P}_{\lambda}\left(u_{\ell}\right)e^{i\left(p_{\ell+1}+...+p_{M+2}\right)}\prod_{k=\ell+1}^{M+1}\frac{[k|p_{k+1}+...+p_{M+2}|\lambda\rangle}{\left\langle k\lambda\right\rangle }.
\end{align}
Substituting the expression for $\mathcal{P}_{\lambda}\left(u_{\ell}\right)$
involving the differential operator $\partial/\partial x_{\,\,\,\dot{A}}^{A}$,
we obtain:
\begin{align}
 & \prod_{k=\ell}^{M+1}\mathcal{P}_{\lambda}\left(u_{k}\right)e^{ip_{M+2}\cdot x}\\
 & =\frac{\left(\bar{u}_{\ell}\right)_{\dot{A}}\lambda^{A}}{\left\langle \ell\lambda\right\rangle }\frac{1}{i}\frac{\partial}{\partial x_{\,\,\,\dot{A}}^{A}}e^{i\left(p_{\ell}+...+p_{M+2}\right)}\prod_{k=\ell+1}^{M+1}\frac{[k|p_{k+1}+...+p_{M+2}|\lambda\rangle}{\left\langle k\lambda\right\rangle }\\
 & =e^{i\left(p_{\ell}+...+p_{M+2}\right)\cdot x}\frac{[\ell|p_{\ell+1}+...+p_{M+2}|\lambda\rangle}{\left\langle \ell\lambda\right\rangle }\prod_{k=\ell+1}^{M+1}\frac{[k|p_{k+1}+...+p_{M+2}|\lambda\rangle}{\left\langle k\lambda\right\rangle }\\
 & =e^{i\left(p_{\ell}+...+p_{M+2}\right)\cdot x}\prod_{k=\ell}^{M+1}\frac{[k|p_{k+1}+...+p_{M+2}|\lambda\rangle}{\left\langle k\lambda\right\rangle }.
\end{align}
This completes the induction proof of Eq. (\ref{eq:Induction-Hypothesis}).

To establish our Proposition, we require the equation above for $\ell=2$
and $m=n-2$, yielding: 
\begin{equation}
\prod_{k=2}^{n-2}\mathcal{P}_{\lambda}\left(u_{k}\right)e^{ip_{n-1}\cdot x}=e^{i\left(p_{2}+...+p_{n-1}\right)\cdot x}\prod_{k=2}^{n-2}\frac{[k|p_{k+1}+...+p_{n-1}|\lambda\rangle}{\left\langle k\lambda\right\rangle }.\label{eq:Id1}
\end{equation}

Using the mode expansions for the fermionic doublet $\chi$ and $\chi^{\dagger}$,
as defined in Eq. (\ref{eq:Modes-2}), the two-point function of these
fields is given by $\langle\chi\left(u_{1}\right)\chi^{\dagger}\left(u_{2}\right)\rangle=1/\left\langle 12\right\rangle $.
Therefore, using Eq. (\ref{eq:Id1}) and contracting the fields $\chi$
and $\chi^{\dagger}$ within the correlation function, we obtain:
\begin{align}
 & \langle\lambda\big|\mathcal{Q}_{1}\prod_{k=2}^{n-2}\mathcal{P}_{k}\mathcal{Q}_{n-1}\mathcal{Q}_{n}\bar{\partial}\big|\lambda\rangle\\
 & =\langle0\big|\chi\left(\lambda\right)\mathcal{Q}_{1}\prod_{k=2}^{n-2}\mathcal{P}_{k}\overbrace{\left[e^{ip_{n-1}}\chi^{\dagger}\left(u_{n-1}\right)\chi\left(u_{n-1}\right)\right]\mathcal{Q}_{n}}^{\mathcal{Q}_{n-1}}\bar{\partial}\chi^{\dagger}\left(\lambda\right)\big|0\rangle\\
 & =\frac{e^{i\left(p_{1}+...+p_{n}\right)\cdot x}}{\left\langle \lambda1\right\rangle \left\langle 1\,\left(n-1\right)\right\rangle \left\langle \left(n-1\right)\,n\right\rangle }\bar{\partial}_{\lambda}\left(\frac{1}{\left\langle n\lambda\right\rangle }\right)\prod_{k=2}^{n-2}\frac{[k|p_{k+1}+...+p_{n}|\lambda\rangle}{\left\langle k\lambda\right\rangle }.\label{eq:Middle-step}
\end{align}

Now, let $z_{\lambda}\in\mathbf{CP}^{1}$ denote the local coordinate
corresponding to the spinor $\lambda^{A}$. Let $\lambda^{A}\eqqcolon\left(\xi_{\lambda},\zeta_{\lambda}\right)^{T}$
be a matrix representation with $\zeta_{\lambda}\neq0$. Then, $z_{\lambda}=\xi_{\lambda}/\zeta_{\lambda}$
in an open neighbourhood $\mathcal{U}\subset\mathbf{CP}^{1}$ such
that $\{\zeta=0\}\cap\mathcal{U}=\emptyset$. Integrating Eq. (\ref{eq:Middle-step})
over $x^{\mu}$ and $z_{\lambda}$, and using the identity $\partial_{\bar{z}}\left(1/z\right)=\pi\delta\left(z\right)$,
while remembering that all four-vectors $p_{1}$, ..., $p_{n}$ are
light-like, we derive a significant result:
\begin{align}
 & \int\left[dxd\lambda\right]\langle\lambda\big|\mathcal{Q}_{1}\prod_{k=2}^{n-2}\mathcal{P}_{k}\mathcal{Q}_{n-1}\mathcal{Q}_{n}\bar{\partial}\big|\lambda\rangle\\
 & =-\frac{1}{\left\langle n\,1\right\rangle \left\langle 1\,\left(n-1\right)\right\rangle \left\langle \left(n-1\right)\,n\right\rangle }\prod_{k=2}^{n-2}\frac{[k|p_{k+1}+...+p_{n-1}|n\rangle}{\left\langle kn\right\rangle }\delta_{P},
\end{align}
where $P=\sum_{k=1}^{n}p_{k}$. 

The final step of our argument hinges on the application of a fundamental
result established by \citet{berends1988relations}. According to
this result, the tree-level MHV $n$-graviton amplitude $M_{n}$ for
$n\geq5$ can be expressed as: 
\begin{align}
M_{n} & =\left\langle 12\right\rangle ^{8}\frac{\left[12\right]\left[\left(n-2\right)\,\left(n-1\right)\right]}{\left\langle 1\,\left(n-1\right)\right\rangle N\left(n\right)}\left(\prod_{i=1}^{n-3}\prod_{j=i+2}^{n-1}\left\langle ij\right\rangle \right)\prod_{\ell=3}^{n-3}[\ell|p_{\ell+1}+...+p_{n-1}|n\rangle\delta_{P}\label{eq:Final-Step-1}\\
 & +\mathcal{P}\left(2,3,...,n-2\right)
\end{align}
where $N\left(n\right)$ is defined as:
\begin{equation}
N\left(n\right)\coloneqq\prod_{i=1}^{n-1}\prod_{j=1+i}^{n}\left\langle ij\right\rangle ,
\end{equation}
and $\mathcal{P}\left(2,3,...,n-2\right)$ denotes permutation of
the indices $\{2,3,...,n-2\}$. 

Given that $N\left(n\right)$ encompasses the product of all terms
$\left\langle ij\right\rangle $ with $j$ ranging from $1+i$ to
$n$ and $i$ from $1$ to $n-1$, while $\prod_{i=1}^{n-3}\prod_{j=i+2}^{n-1}\left\langle ij\right\rangle $
includes only the products with $j$ running from $i+2$ to $n-1$
for $i$ ranging from $1$ to $n-3$, we can express\footnote{The interested reader is encouraged to compute this factor for the
cases $n=3,4,5$ in order to develop an intuitive understanding of
the underlying factorisation. A rigorous proof of this is an easy
exercise in mathematical induction.}: 
\begin{equation}
\frac{1}{N\left(n\right)}\prod_{i=1}^{n-3}\prod_{j=i+2}^{n-1}\left\langle ij\right\rangle =-\frac{1}{C\left(n\right)\left\langle 2\,n\right\rangle \left\langle 3\,n\right\rangle ...\left\langle \left(n-2\right)\,n\right\rangle }.
\end{equation}
Given that we assumed all particles are outgoing, invoking energy-momentum
conservation yields: $[2|p_{3}+...+p_{n-1}|n\rangle=[2|\left(-p_{1}\right)|n\rangle=\left[12\right]\left\langle 1n\right\rangle $,
and additionally, 
\[
[\left(n-2\right)|p_{n-1}|n\rangle=\left[\left(n-2\right)\,\left(n-1\right)\right]\left\langle \left(n-1\right)\,n\right\rangle .
\]
Thus, the expression for $M_{n}$ with $n\geq5$ can be reformulated
as:
\begin{align}
M_{n} & =\frac{\left\langle 12\right\rangle ^{8}}{C\left(n\right)}\frac{1}{\left\langle n\,1\right\rangle \left\langle 1\,\left(n-1\right)\right\rangle \left\langle \left(n-1\right)\,n\right\rangle }\prod_{k=2}^{n-2}\frac{[k|p_{k+1}+...+p_{n-1}|n\rangle}{\left\langle kn\right\rangle }\delta_{P}\label{eq:Final-Step-2}\\
 & +\mathcal{P}\left(2,...,n-2\right).
\end{align}
Finally, Eqs. (\ref{eq:Final-Step-1}, \ref{eq:Final-Step-2}) imply
that:
\begin{equation}
M_{n}=-\frac{\left\langle 12\right\rangle ^{8}}{C\left(n\right)}\int\left[dxd\lambda\right]\langle\lambda\big|\mathcal{Q}_{1}\prod_{k=2}^{n-2}\mathcal{P}_{k}\mathcal{Q}_{n-1}\mathcal{Q}_{n}\bar{\partial}\big|\lambda\rangle+\mathcal{P}\left(2,...,n-2\right),\label{eq:Mn-final}
\end{equation}
for $n\geq5$. 

Let $\theta_{A}^{p}$ ($p=1,...,8$) be Grassmann-valued two-component
spinors satisfying the normalisation condition $\int d^{2}\theta^{p}\,\theta_{A}^{p}\theta_{B}^{p}=\varepsilon_{AB}$
(\citet{berezin2013introduction}). Therefore, the term $\left\langle 12\right\rangle ^{8}$
can be represented in integral form as:
\begin{equation}
\left\langle 12\right\rangle ^{8}=\int d^{16}\theta\,\prod_{p=1}^{8}\theta_{A}^{p}u_{1}^{A}\prod_{q=1}^{8}\theta_{B}^{q}u_{2}^{B}.
\end{equation}
To streamline the notation for our integration measure and to align
it with integrals involving $x$ and $z_{\lambda}$, we introduce
the notation $\int\left[d\theta_{ij}\right]\coloneqq\int d^{16}\theta\,\prod_{p=1}^{8}\theta_{A}^{p}u_{i}^{A}\prod_{q=1}^{8}\theta_{B}^{q}u_{j}^{B}$.
Ultimately, we derive the final result:
\begin{equation}
M_{n}=-\frac{1}{C\left(n\right)}\int\left[d\theta_{12}dxd\lambda\right]\langle\lambda\big|\mathcal{Q}_{1}\prod_{k=2}^{n-2}\mathcal{P}_{k}\mathcal{Q}_{n-1}\mathcal{Q}_{n}\bar{\partial}\big|\lambda\rangle+\mathcal{P}\left(2,3,...,n-2\right).
\end{equation}

\subsection{Frequency Dependence }

It will prove advantageous to re-express Eqs. (\ref{eq:M3-final},
\ref{eq:M4-final}, \ref{eq:Mn-final}) in terms of the frequencies
and the variables associated with the complex projective lines that
parametrise the directions of the graviton momenta. For each $k\geq1$,
the momentum $p_{k}^{\mu}$ of the $k^{\text{th}}$ graviton can be
expressed as $p_{k}^{\mu}=\omega q_{k}^{\mu}$, with $q_{k}^{\mu}$
being the standard null momentum defined by $q_{k}^{\mu}=(\sigma^{\mu})_{A\dot{A}}v_{k}^{A}\bar{v}_{k}^{\dot{A}}$.
Using the little group rescaling property, we may select local coordinates
$z_{k}$, $\bar{z}_{k}$ on an open subset of $\mathbf{CP}^{1}$ such
that $v_{k}^{\dot{A}}=\left(z,1\right)^{T}$ and $\bar{v}_{k}^{\dot{A}}=\left(\bar{z},1\right)^{T}$.
Note that, if we define the two-component spinors $u_{k}^{A}\coloneqq\sqrt[]{\omega_{k}}v_{k}^{A}$
and $\bar{u}_{k}\eqqcolon\sqrt[]{\omega_{k}}\bar{v}_{k}^{\dot{A}}$
in these local coordinates, then $p_{k}^{\mu}=(\sigma^{\mu})_{A\dot{A}}u_{k}^{A}u_{k}^{\dot{A}}$. 

Thus, substituting these components into the aforementioned and letting
$z_{ij}\coloneqq z_{i}-z_{j}$, the three-point amplitude can be written
as:
\begin{equation}
M_{3}=-\frac{\omega_{1}^{3}\omega_{2}^{3}}{\omega_{3}}\frac{z_{12}^{8}}{z_{12}z_{23}z_{31}}\int\left[dxd\lambda\right]\,\langle\lambda\big|\mathcal{Q}_{1}\mathcal{Q}_{2}\mathcal{Q}_{3}\bar{\partial}\big|\lambda\rangle,\label{eq:M3-Frequency}
\end{equation}
while the four-point amplitude becomes:
\begin{equation}
M_{4}=-\frac{\omega_{1}^{3}\omega_{2}^{3}}{\omega_{3}\omega_{4}}\frac{z_{12}^{8}}{z_{12}z_{23}z_{34}z_{41}}\int\left[dxd\lambda\right]\,\langle\lambda\big|\mathcal{Q}_{1}\mathcal{P}_{2}\mathcal{Q}_{3}\mathcal{Q}_{4}\bar{\partial}\big|\lambda\rangle.
\end{equation}
For $n\geq5$, the amplitude generalises to:
\begin{equation}
M_{n}=-\frac{\omega_{1}^{3}\omega_{2}^{3}}{\omega_{3}\omega_{4}...\omega_{n}}\frac{z_{12}^{8}}{z_{12}z_{23}...z_{n1}}\int\left[dxd\lambda\right]\langle\lambda\big|\mathcal{Q}_{1}\prod_{k=2}^{n-2}\mathcal{P}_{k}\mathcal{Q}_{n-1}\mathcal{Q}_{n}\bar{\partial}\big|\lambda\rangle+\mathcal{P}\left(2,3,...,n-2\right).
\end{equation}

\section{Operator Factorisation\label{sec:Operator-Factorisation}}

Let $x$, $y$, $q$, $q_{1}$, $q_{2}$, ... $\in\mathbb{R}^{4}$
be a sequence of four-vectors, and let $\varepsilon$, $\varepsilon_{1}$,
$\varepsilon_{2}$, ... $\in\mathbb{R}$ represent a family of real
variables. For the sake of brevity, the spacetime inner product will
be indicated by simple juxtaposition, $qx\coloneqq q\cdot x$, and
we shall denote $\partial_{\varepsilon}\coloneqq\partial/\partial\varepsilon$.
For any integer $\ell\geq1$ and\footnote{Here, $1+i\mathbb{R}\coloneqq\left\{ 1+i\lambda\,\big|\,\lambda\in\mathbf{R}\right\} $.}
$\Delta\in1+i\mathbb{R}$, let the functions $G_{\Delta}\left(x,q\right)$
and $K_{\ell,\Delta}\left(x,q\right)$ be defined by:
\begin{equation}
G_{\Delta}\left(x,q\right)\coloneqq\frac{\Gamma\left(\Delta\right)}{\left(\varepsilon-iqx\right)^{\Delta}},\,\,\,K_{\ell,\Delta}\left(x,q\right)\coloneqq\frac{\Gamma\left(\Delta\right)}{\left(\varepsilon_{\ell}-iqx\right)^{\Delta}}.
\end{equation}
To simplify notation where no ambiguity arises, we shall write $G_{\ell}=G_{\Delta}\left(x,q_{\ell}\right)$
and $K_{\ell,\Delta}=K_{\ell}\left(x,q_{\ell}\right)$. Note that
the parameters $\varepsilon_{\ell}$ appearing in the functions $K_{\ell,\Delta}\left(x,q\right)$
are independent real variables, while the set of all functions $\{G_{\Delta}\left(x,q\right)\}$
is a one-parameter parametrised $\varepsilon$ for each pair $(\ell$,
$\Delta)$. The form of these functions should already be familiar
to the reader, as they correspond, when restricted to the de Boer-Solodukhin
hyperbolic foliation of Minkowski space (cf. \citet{de2003holographic}),
to the $AdS_{3}$-Witten bulk-to-boundary propagator, modulo the normalisation
constants of the Green function of the Laplacian $\square_{AdS}$.

Now, define the differential operators $\mathbf{X}_{\ell}\left(x\right)$,
$\mathcal{D}_{\ell}\left(y\right)$ and $\mathcal{T}_{\ell}\left(y\right)$
by:
\begin{equation}
\mathbf{X}_{\ell}\left(x\right)\coloneqq M_{\ell}^{A\dot{A}}\frac{\partial}{\partial x^{A\dot{A}}}\frac{\Gamma\left(\Delta_{\ell}\right)}{\left(\varepsilon-iq_{\ell}\cdot x\right)^{\Delta_{\ell}}},
\end{equation}
\begin{equation}
\mathcal{D}_{\ell}\left(y\right)\coloneqq M_{\ell}^{A\dot{A}}\frac{\partial}{\partial y^{A\dot{A}}}e^{-iq_{\ell}\cdot y\frac{\partial}{\partial\varepsilon_{\ell}}},\,\,\,\mathcal{T}_{\ell}\left(y\right)\coloneqq e^{-iq_{\ell}\cdot y\frac{\partial}{\partial\varepsilon_{\ell}}}.
\end{equation}
Then, we wish to prove that:
\begin{equation}
\frac{\partial}{\partial y^{A\dot{A}}}\mathcal{D}_{2}\mathcal{D}_{3}...\mathcal{D}_{n}\mathcal{T}_{n+1}\mathcal{T}_{n+2}\prod_{k=1}^{n+2}K_{\ell}\left(x\right)=\frac{\partial}{\partial x^{A\dot{A}}}\mathcal{D}_{2}\mathcal{D}_{3}...\mathcal{D}_{n}\mathcal{T}_{n+1}\mathcal{T}_{n+2}\prod_{k=1}^{n+2}K_{\ell}\left(x\right)\label{eq:Lemma}
\end{equation}
\begin{equation}
\mathbf{X}_{1}\mathbf{X}_{2}...\mathbf{X}_{n}G_{n+1}G_{n+2}=\int dy\,\delta_{y}\mathcal{D}_{1}\mathcal{D}_{2}...\mathcal{D}_{n}\mathcal{T}_{n+1}\mathcal{T}_{n+2}\prod_{k=1}^{n+2}K_{k}\left(x\right).\label{eq:Theorem}
\end{equation}
For simplicity, we will denote $\int dy\delta_{y}\coloneqq\int d^{4}y\delta^{\left(4\right)}\left(y\right)$
on what follows.

\subsection{Proof of Eq. (\ref{eq:Lemma})}

Start with $n=1$. Then, note that:
\begin{align}
 & \frac{\partial}{\partial y^{A\dot{A}}}\mathcal{D}_{1}\mathcal{T}_{2}\mathcal{T}_{3}\prod_{k=1}^{3}K_{k}\left(x\right)\\
 & =\frac{\partial}{\partial y^{A\dot{A}}}M_{1}^{B\dot{B}}\frac{\partial}{\partial y^{B\dot{B}}}e^{-iq_{1}y\partial_{\varepsilon_{1}}}e^{-iq_{2}y\partial_{\varepsilon_{2}}}e^{-iq_{3}y\partial_{\varepsilon_{3}}}\prod_{k=1}^{3}K_{k}\left(x\right)\\
 & =M_{1}^{B\dot{B}}\frac{\partial}{\partial y^{B\dot{B}}}e^{-iq_{1}y\partial_{\varepsilon_{1}}}e^{-iq_{2}y\partial_{\varepsilon_{2}}}e^{-iq_{3}y\partial_{\varepsilon_{3}}}\sum_{r=1}^{3}\left(-iq_{r}\right)_{A\dot{A}}\frac{\partial}{\partial\varepsilon_{r}}\prod_{k=1}^{3}K_{k}\left(x\right)\\
 & =M_{1}^{B\dot{B}}\frac{\partial}{\partial y^{B\dot{B}}}e^{-iq_{1}y\partial_{\varepsilon_{1}}}e^{-iq_{2}y\partial_{\varepsilon_{2}}}e^{-iq_{3}y\partial_{\varepsilon_{3}}}\frac{\partial}{\partial x^{A\dot{A}}}\prod_{k=1}^{3}K_{k}\left(x\right)\\
 & =\frac{\partial}{\partial x^{A\dot{A}}}\mathcal{D}_{1}\mathcal{T}_{2}\mathcal{T}_{3}\prod_{k=1}^{3}K_{k}\left(x\right).
\end{align}

Now, suppose that, for some $n\geq1$,
\begin{equation}
\frac{\partial}{\partial y^{A\dot{A}}}\mathcal{D}_{2}\mathcal{D}_{3}...\mathcal{D}_{n}\mathcal{T}_{n+1}\mathcal{T}_{n+2}\prod_{k=2}^{n+2}K_{k}\left(x\right)=\frac{\partial}{\partial x^{A\dot{A}}}\mathcal{D}_{2}\mathcal{D}_{3}...\mathcal{D}_{n}\mathcal{T}_{n+1}\mathcal{T}_{n+2}\prod_{k=2}^{n+2}K_{k}\left(x\right).\label{eq:Induction-Hypothesis-1}
\end{equation}
Therefore:
\begin{align}
 & \frac{\partial}{\partial y^{A\dot{A}}}\mathcal{D}_{1}\mathcal{D}_{2}...\mathcal{D}_{n}\mathcal{T}_{n+1}\mathcal{T}_{n+2}\prod_{k=1}^{n+2}K_{k}\left(x\right)\\
 & =\frac{\partial}{\partial y^{A\dot{A}}}\left[\left(M_{1}^{B\dot{B}}\frac{\partial}{\partial y^{B\dot{B}}}e^{-iq_{1}y\partial_{\varepsilon_{1}}}\right)\mathcal{D}_{2}...\mathcal{D}_{n}\mathcal{T}_{n+1}\mathcal{T}_{n+2}\prod_{k=1}^{n+2}K_{k}\left(x\right)\right]\\
 & =M_{1}^{B\dot{B}}\frac{\partial}{\partial y^{B\dot{B}}}e^{-iq_{1}y\partial_{\varepsilon_{1}}}\left(-iq_{1}\right)_{A\dot{A}}\frac{\partial}{\partial\varepsilon_{1}}\left(\mathcal{D}_{2}...\mathcal{D}_{n}\mathcal{T}_{n+1}\mathcal{T}_{n+2}K_{1}\left(x\right)\prod_{k=2}^{n+2}K_{k}\left(x\right)\right)\\
 & +M_{1}^{B\dot{B}}\frac{\partial}{\partial y^{B\dot{B}}}e^{-iq_{1}y\partial_{\varepsilon_{1}}}K_{1}\left(x\right)\frac{\partial}{\partial y^{A\dot{A}}}\left(\mathcal{D}_{2}...\mathcal{D}_{n}\mathcal{T}_{n+1}\mathcal{T}_{n+2}\prod_{k=2}^{n+2}K_{k}\left(x\right)\right).
\end{align}
Using the fact that $\left(-iq_{1}\right)_{A\dot{A}}\partial_{\varepsilon_{1}}$
commutes with $\mathcal{D}_{2}...\mathcal{D}_{n}\mathcal{T}_{n+1}\mathcal{T}_{n+2}\prod_{k=2}^{n+2}K_{k}\left(x\right)$
and applying the induction hypothesis (Eq. (\ref{eq:Induction-Hypothesis-1})),
\begin{align}
 & \frac{\partial}{\partial y^{A\dot{A}}}\mathcal{D}_{1}\mathcal{D}_{2}...\mathcal{D}_{n}\mathcal{T}_{n+1}\mathcal{T}_{n+2}\prod_{k=1}^{n+2}K_{k}\left(x\right)\\
 & =M_{1}^{B\dot{B}}\frac{\partial}{\partial y^{B\dot{B}}}e^{-iq_{1}y\partial_{\varepsilon_{1}}}\left(\left(-iq_{1}\right)_{A\dot{A}}\frac{\partial K_{1}\left(x\right)}{\partial\varepsilon_{1}}\right)\mathcal{D}_{2}...\mathcal{D}_{n}\mathcal{T}_{n+1}\mathcal{T}_{n+2}\left(\prod_{k=2}^{n+2}K_{k}\left(x\right)\right)\\
 & +M_{1}^{B\dot{B}}\frac{\partial}{\partial y^{B\dot{B}}}e^{-iq_{1}y\partial_{\varepsilon_{1}}}K_{1}\left(x\right)\frac{\partial}{\partial x^{A\dot{A}}}\left(\mathcal{D}_{2}...\mathcal{D}_{n}\mathcal{T}_{n+1}\mathcal{T}_{n+2}\prod_{k=2}^{n+2}K_{k}\left(x\right)\right).
\end{align}
Recalling that $\left(iq_{1}\right)_{A\dot{A}}\partial_{\varepsilon_{1}}K_{1}\left(x\right)=\partial K_{1}\left(x\right)/\partial x^{A\dot{A}}$,
the above equation can be rewritten as:
\begin{align}
 & \frac{\partial}{\partial y^{A\dot{A}}}\mathcal{D}_{1}\mathcal{D}_{2}...\mathcal{D}_{n}\mathcal{T}_{n+1}\mathcal{T}_{n+2}\prod_{k=1}^{n+2}K_{k}\left(x\right)\\
 & =\mathcal{D}_{1}\frac{\partial K_{1}\left(x\right)}{\partial x^{A\dot{A}}}\mathcal{D}_{2}...\mathcal{D}_{n}\mathcal{T}_{n+1}\mathcal{T}_{n+2}\left(\prod_{k=2}^{n+2}K_{k}\left(x\right)\right)\\
 & +\mathcal{D}_{1}K_{1}K_{1}\left(x\right)\frac{\partial}{\partial x^{A\dot{A}}}\left(\mathcal{D}_{2}...\mathcal{D}_{n}\mathcal{T}_{n+1}\mathcal{T}_{n+2}\prod_{k=2}^{n+2}K_{k}\left(x\right)\right)\\
 & =\frac{\partial}{\partial x^{A\dot{A}}}\mathcal{D}_{1}\mathcal{D}_{2}...\mathcal{D}_{n}\mathcal{T}_{n+1}\mathcal{T}_{n+2}\prod_{k=1}^{n+2}K_{k}\left(x\right),
\end{align}
completing the proof of Eq. (\ref{eq:Lemma}) by induction.

\subsection{Proof of Eq. (\ref{eq:Theorem})\label{subsec:Proof-of-Eq.}}

Let's start with $n=1$,
\begin{align}
\mathbf{X}_{1}G_{2}G_{3} & =M_{1}^{A\dot{A}}\frac{\partial}{\partial x^{A\dot{A}}}\left[\frac{\Gamma\left(\Delta_{1}\right)}{\left(\varepsilon-iq_{1}x\right)^{\Delta_{1}}}\frac{\Gamma\left(\Delta_{1}\right)}{\left(\varepsilon-iq_{1}x\right)^{\Delta_{1}}}\frac{\Gamma\left(\Delta_{1}\right)}{\left(\varepsilon-iq_{1}x\right)^{\Delta_{1}}}\right]\\
 & =-M_{1}^{A\dot{A}}\sum_{k=1}^{3}\left(-iq_{k}\right)_{A\dot{A}}\prod_{\ell=1}^{3}\frac{\Gamma\left(\Delta_{\ell}+\delta_{k\ell}\right)}{\left(\varepsilon-iq_{\ell}x\right)^{\Delta_{\ell}+\delta_{k\ell}}}.
\end{align}
Because:
\[
\int dy\delta_{y}\frac{\partial}{\partial y^{A\dot{A}}}e^{-iqy\partial_{\varepsilon_{\ell}}}\frac{\Gamma\left(\Delta_{\ell}\right)}{\left(\varepsilon_{\ell}-iq_{\ell}x\right)^{\Delta_{\ell}}}=\frac{\partial}{\partial x^{A\dot{A}}}\frac{\Gamma\left(\Delta_{\ell}\right)}{\left(\varepsilon_{\ell}-iq_{\ell}x\right)^{\Delta_{\ell}}},
\]
we have:
\begin{align}
\mathbf{X}_{1}G_{2}G_{3} & =M_{1}^{A\dot{A}}\int dy\,\delta_{y}\frac{\partial}{\partial y^{A\dot{A}}}\left(\prod_{k=1}^{3}e^{-iq_{k}y\partial_{\varepsilon_{k}}}\right)\prod_{\ell=1}^{3}\frac{\Gamma\left(\Delta_{\ell}\right)}{\left(\varepsilon_{\ell}-iq_{\ell}x\right)^{\Delta_{\ell}}}\\
 & =\int dy\,\delta_{y}\mathcal{D}_{1}\mathcal{T}_{2}\mathcal{T}_{3}\prod_{k=1}^{3}K_{k}\left(x\right).
\end{align}

Now, suppose that, for some $n\geq1$,
\begin{equation}
\mathbf{X}_{2}\mathbf{X}_{3}...\mathbf{X}_{n}G_{n+1}G_{n+2}=\int dy\delta_{y}\mathcal{D}_{2}\mathcal{D}_{3}...\mathcal{D}_{n}\mathcal{T}_{n+1}\mathcal{T}_{n+2}\prod_{k=2}^{n+2}K_{k}\left(x\right).\label{eq:Induction-Hypothesis-2}
\end{equation}
Thus:
\begin{align}
 & \int dy\delta_{y}\mathcal{D}_{1}\mathcal{D}_{2}...\mathcal{D}_{n}\mathcal{T}_{n+1}\mathcal{T}_{n+2}\prod_{k=1}^{n+2}K_{k}\left(x\right)\\
 & =\int dy\delta_{y}\left(M_{1}^{A\dot{A}}\frac{\partial}{\partial y^{A\dot{A}}}e^{-iq_{1}y\partial_{\varepsilon_{1}}}\right)\mathcal{D}_{2}...\mathcal{D}_{n}\mathcal{T}_{n+1}\mathcal{T}_{n+2}\frac{\Gamma\left(\Delta_{1}\right)}{\left(\varepsilon_{1}-iq_{1}x\right)^{\Delta_{1}}}\prod_{k=2}^{n+2}K_{k}\left(x\right)\\
 & =\int dy\delta_{y}M_{1}^{A\dot{A}}\frac{\partial}{\partial y^{A\dot{A}}}\left(e^{-iq_{1}y\partial_{\varepsilon_{1}}}\frac{\Gamma\left(\Delta_{1}\right)}{\left(\varepsilon_{1}-iq_{1}x\right)^{\Delta_{1}}}\mathcal{D}_{2}...\mathcal{D}_{n}\mathcal{T}_{n+1}\mathcal{T}_{n+2}\prod_{k=2}^{n+2}K_{k}\left(x\right)\right)\\
 & =M_{1}^{A\dot{A}}\int dy\delta_{y}\left(-iq_{1}\right)_{A\dot{A}}\left(\frac{\partial}{\partial\varepsilon_{1}}\frac{\Gamma\left(\Delta_{1}\right)}{\left(\varepsilon_{1}-iq_{1}x\right)^{\Delta_{1}}}\right)\mathcal{D}_{2}...\mathcal{D}_{n}\mathcal{T}_{n+1}\mathcal{T}_{n+2}\prod_{k=2}^{n+2}K_{k}\left(x\right)\\
 & +M_{1}^{A\dot{A}}\frac{\Gamma\left(\Delta_{1}\right)}{\left(\varepsilon_{1}-iq_{1}x\right)^{\Delta_{1}}}\int dy\delta_{y}\frac{\partial}{\partial y^{A\dot{A}}}\left(\mathcal{D}_{2}...\mathcal{D}_{n}\mathcal{T}_{n+1}\mathcal{T}_{n+2}\prod_{k=2}^{n+2}K_{k}\left(x\right)\right)
\end{align}
Finally, noting that:
\begin{equation}
\left(-iq_{1}\right)_{A\dot{A}}\frac{\partial}{\partial\varepsilon_{1}}\frac{\Gamma\left(\Delta_{1}\right)}{\left(\varepsilon_{1}-iq_{1}x\right)^{\Delta_{1}}}=\frac{\partial}{\partial x^{A\dot{A}}}\frac{\Gamma\left(\Delta_{1}\right)}{\left(\varepsilon_{1}-iq_{1}x\right)^{\Delta_{1}}},
\end{equation}
using Eq. (\ref{eq:Lemma}), and applying the induction hypothesis
(Eq. (\ref{eq:Induction-Hypothesis})), we derive:
\begin{align*}
 & \int dy\delta_{y}\mathcal{D}_{1}\mathcal{D}_{2}...\mathcal{D}_{n}\mathcal{T}_{n+1}\mathcal{T}_{n+2}\prod_{k=1}^{n+2}K_{k}\left(x\right)\\
 & =M_{1}^{A\dot{A}}\left(\frac{\partial}{\partial x^{A\dot{A}}}\frac{\Gamma\left(\Delta_{1}\right)}{\left(\varepsilon_{1}-iq_{1}x\right)^{\Delta_{1}}}\right)\int dy\delta_{y}\mathcal{D}_{2}...\mathcal{D}_{n}\mathcal{T}_{n+1}\mathcal{T}_{n+2}\prod_{k=2}^{n+2}K_{k}\left(x\right)\\
 & +M_{1}^{A\dot{A}}\frac{\Gamma\left(\Delta_{1}\right)}{\left(\varepsilon_{1}-iq_{1}x\right)^{\Delta_{1}}}\frac{\partial}{\partial x^{A\dot{A}}}\int dy\delta_{y}\mathcal{D}_{2}...\mathcal{D}_{n}\mathcal{T}_{n+1}\mathcal{T}_{n+2}\prod_{k=2}^{n+2}K_{k}\left(x\right)\\
 & =\underbrace{M_{1}^{A\dot{A}}\frac{\partial}{\partial x^{A\dot{A}}}\frac{\Gamma\left(\Delta_{1}\right)}{\left(\varepsilon_{1}-iq_{1}x\right)^{\Delta_{1}}}}_{\mathbf{X}_{1}}\mathbf{X}_{2}\mathbf{X}_{3}...\mathbf{X}_{n}G_{n+1}G_{n+2}\\
 & =\mathbf{X}_{1}\mathbf{X}_{2}...\mathbf{X}_{n}G_{n+1}G_{n+2},
\end{align*}
completing our inductive proof.

\bibliographystyle{aapmrev4-2}
\bibliography{GravityLiouville}

\end{document}